\documentclass[superscriptaddress,groupedaddress,nofootnoteinbib,11pt]{article}
\pdfoutput=1 
\usepackage{setspace}
\usepackage{jheppub}

\usepackage[utf8x]{inputenc}
\usepackage{mathtools,slashed,mathrsfs}
\usepackage[caption=false]{subfig}
\usepackage{dcolumn}
\usepackage{multirow}
\usepackage{tabularx}
\usepackage{booktabs}
\usepackage{bm}
\usepackage{comment}
\usepackage[dvipsnames]{xcolor}
\usepackage[normalem]{ulem} 
\usepackage{enumerate}
\usepackage{siunitx}
\usepackage{xspace}
\usepackage{adjustbox}
\allowdisplaybreaks

\newcommand{\Sec}[1]{Sec.~\ref{#1}}

\newcommand{\Fig}[1]{Fig.~\ref{#1}}

\newcommand{\beq}{\begin{equation}}
\newcommand{\eeq}{\end{equation}}
\newcommand{\ba}{\begin{array}}
\newcommand{\ea}{\end{array}}
\newcommand{\bea}{\begin{eqnarray}}
\newcommand{\eea}{\end{eqnarray} }
\newcommand{\be}{\begin{eqnarray}}
\newcommand{\ee}{\end{eqnarray}}
\newcommand{\bal}{\begin{align}}
\newcommand{\eal}{\end{align}}
\newcommand{\bi}{\begin{itemize}}
\newcommand{\ei}{\end{itemize}}
\newcommand{\ben}{\begin{enumerate}}
\newcommand{\een}{\end{enumerate}}
\newcommand{\bc}{\begin{center}}
\newcommand{\ec}{\end{center}}
\newcommand{\bt}{\begin{table}}
\newcommand{\et}{\end{table}}
\newcommand{\btb}{\begin{tabular}}
\newcommand{\etb}{\end{tabular}}

\newcommand{\bl}{\left}
\newcommand{\br}{\right}
\newcommand{\eg}{\textit{e.g.}}

\newcommand{\MeV}{\mathrm{MeV}}

\newcommand{\Mpc}{\mathrm{Mpc}}

\newcommand{\km}{\mathrm{km}}
\newcommand{\seg}{\mathrm{s}}

\newcommand{\mA}{\mathcal{A}}
\newcommand{\mH}{\mathcal{H}}
\newcommand{\mF}{\mathcal{F}}

\newcommand{\mAH}{\mathcal{AH}}

\newcommand{\mP}{\mathcal{P}}
\newcommand{\mPH}{\mathcal{PH}}

\newcommand{\mPF}{\mathcal{PF}}
\newcommand{\mPHF}{\mathcal{PHF}}

\newcommand{\eq}{\ensuremath{\mathrm{eq}}}
\newcommand{\dec}{\ensuremath{\mathrm{dec}}}

\newcommand{\tr}{\mathrm{tr}}

\newcommand{\lcdm}{\ensuremath{\Lambda\mathrm{CDM}}\xspace}

\newcommand{\dm}{\mathrm{dm}}

\newcommand{\idm}{\mathrm{idm}}

\newcommand{\dr}{\mathrm{dr}}

\newcommand{\logzdec}{\ensuremath{\log_{10}\bl( z_\mathrm{dec} \br)}}
\newcommand{\logztr}{\ensuremath{\log_{10}\bl( z_\mathrm{tr} \br)}}

\newcommand{\Neff}{\ensuremath{N_{\mathrm{eff}}}\xspace}
\newcommand{\DNeff}{\ensuremath{\Delta N_\mathrm{eff}}\xspace}

\newcommand{\HO}{\ensuremath{H_0}\xspace}

\newcommand{\genidm}{\ensuremath{\text{gen-iDM}}\xspace}

\global\long\def\b#1{\left(#1\right)}%

\definecolor{bleudefrance}{rgb}{0.19, 0.55, 0.91}

\newcommand{\ignore}[1]{}

\usepackage{tikz,xcolor,hyperref}
\definecolor{lime}{HTML}{A6CE39}
\DeclareRobustCommand{\orcidicon}{%
	\begin{tikzpicture}
	\draw[lime, fill=lime] (0,0) 
	circle [radius=0.16] 
	node[white] {{\fontfamily{qag}\selectfont \tiny ID}};	\draw[white, fill=white] (-0.0625,0.095) 
	circle [radius=0.007];	\end{tikzpicture}
	\hspace{-2mm}}
\foreach \x in {A,...,Z}{%
	\expandafter\xdef\csname orcid\x\endcsname{\noexpand\href{https://orcid.org/\csname orcidauthor\x\endcsname}{\noexpand\orcidicon}}
}

\begin{document}
\preprint{\begin{flushright}
    UT-WI-40-2025\\
\end{flushright}}

\title{Dark Matter-Dark Radiation Interactions and the Hubble Tension}

\date{\today}
\author[a,b]{Manuel A. Buen-Abad,\orcidA{}}
\author[a]{Zackaria Chacko,\orcidB{}}
\author[a]{Ina Flood,\orcidC{}}
\author[c]{Can Kilic,\orcidD{}}
\author[d]{Gustavo Marques-Tavares,\orcidE{}}
\author[e,f]{Taewook Youn,\orcidF{}}

\affiliation[a]{Maryland Center for Fundamental Physics, Department of Physics, University of Maryland, College Park, MD 20742, U.S.A.}
\affiliation[b]{Dual CP Institute of High Energy Physics, C.P. 28045, Colima, M\'{e}xico}
\affiliation[c]{Theory Group, Weinberg Institute for Theoretical Physics, University of Texas at Austin, Austin, TX 78712, U.S.A.}
\affiliation[d]{Department of Physics and Astronomy, University of Utah, Salt Lake City, UT 84112, U.S.A.}
\affiliation[e]{Laboratory for Elementary Particle Physics, Cornell University, Ithaca, NY 14853, USA}
\affiliation[f]{School of Physics, Korea Institute for Advanced Study,  Seoul 02455, Republic of Korea}
\emailAdd{buenabad@umd.edu}
\emailAdd{zchacko@umd.edu}
\emailAdd{iflood@umd.edu}
\emailAdd{kilic@physics.utexas.edu}
\emailAdd{g.marques@utah.edu}
\emailAdd{taewook.youn@cornell.edu}

\abstract{

Models in which a subcomponent of dark matter interacts with dark radiation have been proposed as a solution to the Hubble tension. In this framework, the interacting subcomponent of dark matter is in thermal equilibrium with the dark radiation in the early universe, but decouples from it around the time of matter-radiation equality. We study this general class of models and evaluate the quality of fit to recent cosmological data on
the cosmic microwave background (from Planck 2018 and ACT DR6), baryon acoustic oscillations, large-scale structure, supernovae type Ia, and Cepheid variables. We focus on three benchmark scenarios that differ in the rate at which the dark matter decouples from the dark radiation,
resulting in different patterns of dark acoustic oscillations. Fitting without ACT DR6 data, we find that all three scenarios significantly reduce the Hubble tension relative to $\Lambda$CDM, with an exponentially fast decoupling being the most preferred. The tension is reduced to less than $2 \, \sigma$ in fits that don't include the SH0ES collaboration results as part of the data and to less than $1 \, \sigma$ when these are included. The preference for the interacting model is substantially reduced when ACT DR6 data are included. We find that the largest $H_0$ value at the $95 \%$ confidence region is $70.1$ km/s/Mpc without the SH0ES data, leading to only a mild reduction in the tension. This increases to $72.5$ km/s/Mpc, corresponding to a reduction in the tension to less than $3 \, \sigma$, if the SH0ES results are included in the fit.
}

\maketitle

\section{Introduction}
\label{sec:intro}

Despite the general success of the \lcdm model in describing cosmological data, the model faces several challenges. Possibly the most serious of these challenges is the persistent tension between different measurements of \HO, the current expansion rate of the universe. Two of the most precise measurements for \HO are the \lcdm fit to Planck cosmic microwave background (CMB) data~\cite{Planck:2018vyg}, which gives \HO = \SI[separate-uncertainty=true,multi-part-units=single]{67.36 \pm 0.54}{\km/s/\Mpc}, and supernovae observations made by the SH0ES collaboration using Cepheid variable stars to establish a distance ladder~\cite{Riess:2021jrx}, which give \HO = \SI[separate-uncertainty=true,multi-part-units=single]{73.04 \pm 1.04}{\km/s/\Mpc}. It has been proposed that the tension between these two measurements, which exceeds 5$\sigma$, could stem from some unknown systematics in the supernovae measurements~\cite{Freedman:2021ahq,Kamionkowski:2022pkx}. However, the distance ladder results from the SH0ES collaboration have remained consistent through variations in the data used and changes to the analysis methods, and have withstood significant scrutiny\footnote{It is worth noting that the Chicago-Carnegie Hubble Program find slightly smaller values for $H_0$~\cite{Freedman:2024eph}, which are in less tension with the CMB measurement, albeit with larger uncertainty.}. The search for a resolution to this \HO tension has inspired a broad class of models beyond \lcdm (see, \eg, Refs.~\cite{Buen-Abad:2015ova,Lesgourgues:2015wza,Buen-Abad:2017gxg,Zhao:2017cud,DiValentino:2017gzb,Poulin:2018cxd,Smith:2019ihp,Lin:2019qug,Alexander:2019rsc,Agrawal:2019lmo,Escudero:2019gvw,Berghaus:2019cls,Vagnozzi:2019ezj,Ye:2020btb,Das:2020xke,RoyChoudhury:2020dmd,Brinckmann:2020bcn,Krishnan:2020vaf,Seto:2021xua,Johnson:2021wou,Ye:2021iwa,Niedermann:2021vgd,Aloni:2021eaq, Dainotti:2021pqg,Odintsov:2022eqm,Berghaus:2022cwf,Schoneberg:2022grr,Joseph:2022jsf,Buen-Abad:2022kgf,Rezazadeh:2022lsf,Brinckmann:2022ajr,Wang:2022bmk,Bansal:2022qbi,Buen-Abad:2023uva,Sandner:2023ptm,Niedermann:2023ssr,Hughes:2023tcn,Greene:2024qis,Allali:2024anb, Co:2024oek,Cho:2024lhp,Simon:2024jmu,DeSimone:2024lvy,Montani:2024ntj,Buen-Abad:2024tlb,Garny:2025kqj,Bansal:2025usn,GarciaEscudero:2025orc,Chang:2025uvx} for a partial list of these proposals, and also the models reviewed in~\cite{DiValentino:2021izs,Schoneberg:2021qvd, Abdalla:2022yfr,Escudero:2022rbq,Poulin:2023lkg,Khalife:2023qbu,CosmoVerseNetwork:2025alb}).

A promising class of proposed solutions to the Hubble tension are those that only modify the \lcdm paradigm at early times, during the CMB epoch~\cite{Poulin:2024ken}. New physics at these early times can alter the sound horizon scale, thereby modifying the value of \HO inferred from CMB data~(see Refs.~\cite{Buen-Abad:2017gxg,Poulin:2018cxd,Smith:2019ihp,Lin:2019qug,Alexander:2019rsc,Agrawal:2019lmo,Escudero:2019gvw,Berghaus:2019cls,Ye:2020btb,Das:2020xke,RoyChoudhury:2020dmd,Brinckmann:2020bcn,Seto:2021xua,Ye:2021iwa,Niedermann:2021vgd,Aloni:2021eaq,Berghaus:2022cwf,Schoneberg:2022grr,Joseph:2022jsf,Buen-Abad:2022kgf,Rezazadeh:2022lsf,Brinckmann:2022ajr,Bansal:2022qbi,Buen-Abad:2023uva,Sandner:2023ptm,Hughes:2023tcn,Greene:2024qis,Allali:2024anb, Co:2024oek,Cho:2024lhp,Simon:2024jmu,Buen-Abad:2024tlb,Chang:2025uvx,Garny:2025kqj,GarciaEscudero:2025orc} for a partial list of proposals on how to address the Hubble tension by changing the sound horizon). Specifically, because the ratio of the sound horizon scale and the angular distance to recombination must remain fixed to reproduce the locations of the CMB peaks, the contributions of new physics to the energy density around recombination, which would directly decrease the sound horizon scale, must be compensated for by an increase in the energy density at low redshifts that decreases the angular diameter distance, thereby increasing the value of \HO.
Since the model at later times is just \lcdm, this class of proposed solutions has the potential to alleviate the Hubble tension without significantly affecting the fit to late-time observations from baryon acoustic oscillation (BAO) measurements~\cite{Beutler:2011hx,Ross:2014qpa,eBOSS:2020yzd,eBOSS:2020qek,eBOSS:2020fvk,eBOSS:2020gbb,eBOSS:2020uxp,eBOSS:2020tmo} and Type Ia supernova luminosities~\cite{Scolnic:2021amr,Brout:2022vxf}.


A simple mechanism to increase the energy density at recombination involves the introduction of new states in the dark sector. The most basic of these scenarios simply introduces a massless degree of freedom that takes the form of dark radiation (DR) during the CMB epoch, with an abundance parametrized by $\DNeff$. However, this simple proposal is disfavored by the data; the change in the Hubble scale affects the sound horizon scale and the diffusion damping scales differently, leading to an increase in Silk damping at small scales~\cite{Blinov:2020hmc,Hou:2011ec}. This challenge motivates exploring more complex dark sector models. A non-minimal dark sector could contain interactions between dark matter (DM) and the DR. Interactions of DM with DR have the effect of suppressing structure growth until the interaction freezes out. This gives rise to a large suppression of the matter power spectrum (MPS) for modes that entered the horizon prior to when the interaction froze out~\cite{Ackerman:2008kmp,Cyr-Racine:2015ihg}. Such a feature would be highly distinctive and is severely constrained by Lyman alpha data and observations of small dwarf galaxies~\cite{Boehm:2001hm,Cyr-Racine:2013fsa,Schewtschenko:2014fca,Irsic:2017ixq,Archidiacono:2019wdp,DES:2020fxi}. However, if only a subcomponent of the DM participates in the interactions with the DR, these constraints become much weaker. In particular, it has been recently shown that interacting dark sectors in which $\mathcal{O}(1\%)$ of the DM density is coupled to DR, and the interaction freezes out close to matter-radiation equality, can significantly increase the allowed range for the DR density. This allows for a larger value of the extracted value of $H_0$~\cite{Buen-Abad:2024tlb}, opening up a promising avenue to address the tension.

A specific model that realizes this scenario was studied in Ref.~\cite{Buen-Abad:2024tlb}. In this construction, the interacting DM (iDM) component consists of dark electrons and dark protons that undergo recombination into dark atoms once the dark sector temperature is sufficiently low. In such a framework, the interaction rate between DM and DR evolves in a manner similar to the baryon-photon interactions in the Standard Model (SM), and becomes exponentially suppressed with redshift once dark recombination occurs. In this paper, in order to further explore the potential of non-minimal dark sectors to address the Hubble tension, we explore a more general set of interactions between DR and DM, which we call generalized iDM (\genidm). Following an approach similar to that of Ref.~\cite{Cyr-Racine:2015ihg}, we consider different possibilities for the evolution with temperature of the momentum exchange rate between DM and DR, $\Gamma_d$. This affects the rate at which the iDM and DR decouple, impacting cosmological observables. We focus on three benchmark scenarios in which $\Gamma_d$ scales as $T^{2+n}$, where $n=4$, $n=2$, and $n=0$, and determine how well each case addresses the Hubble tension.
While these three benchmarks were chosen for phenomenological clarity, each of them maps into one or more well-motivated physical scenarios. 
For each case, we evaluate the quality of fit to recent cosmological data from the CMB, including the recently released ACT DR6 power-spectrum likelihood~\cite{AtacamaCosmologyTelescope:2025nti,AtacamaCosmologyTelescope:2025blo}, BAO, large-scale structure, Type Ia supernovae, and Cepheid variables. To the best of our knowledge, this provides the first assessment of the impact of the ACT DR6 power-spectrum data on the class of interacting dark-sector models considered here.
We find that all three scenarios reduce the Hubble tension relative
to $\lcdm$, with an exponentially fast decoupling being the most preferred.

This paper is organized as follows. In Sec.~\ref{sec:model}, we describe the framework we are considering and present the three benchmark cases, along with examples of physical models that realize them. In Sec.~\ref{sec:meth} we discuss the methodology and data sets used for our numerical analysis. In Sec.~\ref{sec:results}, we analyze the results of the MCMC fits for each case. We conclude in Sec.~\ref{sec:conclusions}. Our full numerical results can be found in App.~\ref{appendix}.

\section{The Framework}
\label{sec:model}

We consider a framework in which, while the primary component of DM is cold and non-interacting, a subcomponent of DM is in thermal equilibrium with DR at early times. This iDM subcomponent eventually decouples from the DR during the CMB epoch. We assume that the DR is populated only after Big Bang nucleosynthesis (BBN), at temperatures $T_\gamma < 0.1~\MeV$. See Refs.~\cite{Aloni:2023tff,Garny:2025kqj} for possible mechanisms to generate $\DNeff$ after BBN. For concreteness, the DR is taken to be always self-interacting, even after decoupling from the iDM.

The equations describing the evolution of perturbations for the DR and the iDM component, in the conformal Newtonian gauge, take the form~\cite{Ma:1995ey,Cyr-Racine:2012tfp,Cyr-Racine:2013fsa,Bansal:2021dfh,Bansal:2022qbi,Hughes:2023tcn}:
\bea
    \dot{\delta}_\idm & = & -\theta_\idm + 3 \dot\phi \ , \\
    \dot{\theta}_\idm & = & - \mathcal{H} \theta + c_{d,s}^2 k^2 \delta_\idm + k^2 \psi + S \gamma \Gamma_d \bl( \theta_\dr - \theta_\idm \br) \ , \\
    \dot{\delta}_\dr & = & - \frac{4}{3} \theta_\dr + 4 \dot\phi \ ,\\
    \dot{\theta}_\dr & = & \frac{1}{4}\delta_\dr + k^2 \psi + \gamma \Gamma_d \bl( \theta_\idm - \theta_\dr \br) \ .
\eea
Here the dots denote derivatives with respect to the conformal time $\tau$, $\mathcal{H} \equiv \dot{a}/a$ is the conformal Hubble rate, $\phi$ and $\psi$ are the gravitational potentials in the spacetime metric, $k$ is the wavenumber of the perturbation in question, and $S \equiv 4 \rho_\dr / 3 \rho_\idm$. The coupling between DR and the iDM component is incorporated into the momentum-exchange rate~$\Gamma_d$. 

Following an approach analogous to that in Ref.~\cite{Cyr-Racine:2015ihg}, we consider phenomenological scenarios in which $\Gamma_d$ is a simple function of temperature of the form,
\begin{equation}
    \Gamma_d(T) = \Gamma_d^0 \left( \frac{T}{T_0}\right)^{2+n} \, .
\end{equation}
 Many well-motivated models of DM-DR interactions give rise to a momentum exchange rate that evolves with temperature in this simple way. We will focus our analysis on benchmark cases corresponding to $n=4, \, n=2$, and $n=0$, and investigate the impact of the rate of decoupling on cosmological observables. In the $n=0$ case, the interaction would not decouple during radiation domination since it has the same temperature scaling as the Hubble rate, so we consider a scenario in which the interaction rate takes this form at early times but becomes exponentially suppressed at some specific redshift. This is similar to the dynamics in Refs.~\cite{Cyr-Racine:2012tfp,Buen-Abad:2022kgf,Garny:2025kqj}. In all three scenarios, the impact on cosmological observables is effectively determined by three parameters, the amount of dark radiation (\DNeff), the fraction of the DM energy density in iDM, parameterized in terms of $f_\idm = \rho_\idm/\rho_\text{dm}$, and the redshift $z_\text{dec}$ at which the iDM-iDR interactions decouple. In the subsections below, we briefly describe examples of models that realize each of these benchmark cases.

\begin{figure}[tb]
	\centering
	\includegraphics[width=.95\linewidth]{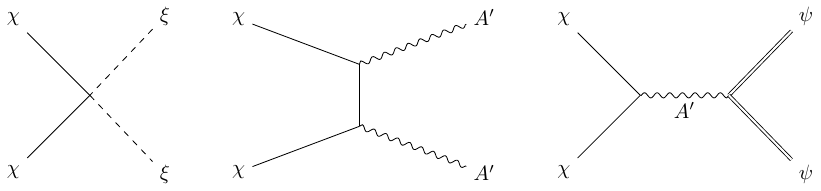}
	\caption{Feynman diagrams for different dark sector interactions considered in this work. $\chi$ denotes the interacting dark matter (iDM), while $\xi$, $A'$, and $\psi$ represent the dark radiation (DR) components corresponding to $n=4$, $n=2$, and $n=0$ cases, respectively.
    {\it Left:} Four-fermion interaction ($n=4$).
    {\it Middle:} Compton-like scattering ($n=2$).
    {\it Right:} Coulomb-like scattering  ($n=0$).
 }
	\label{fig:diagrams}
\end{figure}

\subsection{$n=4$ : Four-Fermion Interaction}

This case arises if the iDM subcomponent is composed of a massive Dirac fermion $\chi$ that interacts through a coupling of the four-fermion form with the DR that is composed of massless fermions $\xi$,
\begin{equation}
    \delta \mathcal{L} = G_d \bar \chi \gamma^\mu \chi \bar \xi \gamma_\mu \xi \, .
\end{equation}
An effective operator of this form can arise in models in which a vector mediator that is heavy compared to the momentum exchange in the collision has been integrated out. It is somewhat reminiscent of the Fermi interactions that describe the weak decays at low energies. 

The iDM and DR are kept in thermal equilibrium at early times through the Feynman diagram shown in the left panel of \Fig{fig:diagrams}. For $T \ll M_\chi$, the corresponding momentum transfer rate scales as
\begin{equation}
    \Gamma_d \sim n_{DR}\sigma_{\xi\chi}v_{\text{rel}}\left(\frac{\delta p}{p_{DM}}\right)^2 \, .
\end{equation}
In this expression, $n_{DR} \sim T^3/\pi^2$ is the number density of the DR, $\sigma_{\xi\chi} \sim G_d^2 T^2/\pi^2$ is the cross section of the four-fermion interaction, $v_{\text{rel}} = 1$ is the relative velocity between the iDM and DR, $\delta p \sim T$ is the change in momentum from a single collision, and $p_{DM} \sim \sqrt{M_\chi T}$ is the momentum of the iDM particles. The $\delta p/p_{DM}$ term arises because the momentum transfer rate is determined by the collision rate divided by the number of collisions necessary for an order one fraction of the momentum to be transferred. This term is squared because the momentum transfer occurs as a random walk~(see, e.g., \cite{Cyr-Racine:2015ihg,Buen-Abad:2017gxg} for a detailed derivation). Therefore, the momentum transfer rate scales as
\begin{equation}
    \Gamma_d \sim  \frac{G_d^2}{\pi^4 M_\chi} T^6\, ,
\end{equation}
which corresponds to $n=4$.

\subsection{$n=2$ : Compton-like scattering}

This case is inspired by the Compton scattering between electrons and photons in the SM. Accordingly, we consider a gauge interaction between a massive Dirac fermion $\chi$, which constitutes the iDM,  and a massless gauge boson $A'$, which constitutes the DR,
\begin{equation}
    \delta \mathcal{L} = g'\bar{\chi} \gamma^{\mu} A'_{\mu} \chi\, .
\end{equation}
The iDM and DR are kept in thermal equilibrium at early times through the Feynman diagram shown in the center panel of \Fig{fig:diagrams}.
As in the $n=4$ case, when $T \ll M_{\chi}$ the momentum transfer rate scales as
\begin{equation}
    \Gamma_d \sim n_{\text{DR}}\, \sigma_{C}\, v_{\text{rel}}\left(\frac{\delta p}{p_{DM}}\right)^2\,
\end{equation}
 where $\sigma_{C} \sim \pi\alpha_d^2/M_{\text{iDM}}^2$ is now the standard Thompson scattering cross section, with $\alpha_d=g'^2/4\pi$. Then the momentum transfer rate scales as, 
\begin{equation}
    \Gamma_d \sim \frac{\alpha_d^2}{\pi M^3_{\chi}}T^4\, ,
\end{equation}
which corresponds to $n=2$. Models in which the DR is composed of a massless scalar that has a Yukawa coupling to a Dirac fermion that constitutes iDM would also give rise to the same scaling behavior.

\subsection{$n=0$ : Coulomb-like scattering}


This corresponds to the case when the momentum transfer rate $\Gamma_d$ between the iDM and the DR scales with temperature as $T^2$ at early times, but falls away exponentially quickly below some specific redshift, so that at late times  
\begin{equation}
    \Gamma_d \propto T^2 \, \exp{\b{-\frac{a}{a_\tr}}} \, .
\end{equation}
To see how this scenario can arise, consider the setup proposed in the Stepped Partially Acoustic DM model (SPartAcous) of Ref.~\cite{Buen-Abad:2023uva}. Here the iDM is composed of a complex scalar $\chi$ that is charged under a dark $U(1)$ gauge symmetry. The gauge boson $A'_\mu$ of the dark gauge symmetry is one of the components of the DR, along with several other massless fields. In addition, the theory contains a light Dirac fermion $\psi$ that also carries charge under the gauge symmetry and has mass $m_\psi$ around the eV scale. The relevant part of the interaction Lagrangian takes the form 
\begin{equation}
    \delta \mathcal{L} = g_d \bar{\psi}\gamma^\mu A_\mu'\psi + i g_d A_\mu' (\chi^\dagger\partial^\mu\chi - \partial_\mu \chi^\dagger\chi).
\end{equation}
The fermions $\psi$ behave like DR at early times but annihilate away into dark gauge bosons once the temperature falls below their mass. The iDM and the fermions $\psi_i$ are kept in thermal equilibrium at early times through the Feynman diagram shown in the right panel of \Fig{fig:diagrams}.
The corresponding momentum transfer rate is given by
\begin{equation}
    \Gamma_d \sim n_{\psi}\sigma_{\chi \psi} v_{\text{rel}}\left(\frac{\delta p}{p_{DM}}\right)^2\,
\end{equation}
where now $\sigma_{\chi \psi} \sim \pi \alpha_d^2/T^2$ and $\alpha_d = g_d^2/4\pi$. Then, for $T > M_\psi$, the momentum exchange rate scales as
\begin{equation}
    \Gamma_d \sim \frac{\alpha_d^2}{\pi M_\chi} T^2\, .
\end{equation}
Once the temperature drops below the mass $m_\psi$ of the massive DR component $\psi$, it departs from the thermal bath, resulting in an exponential suppression factor $\exp{\b{-{m_\psi}/{T}}} \simeq \exp{\b{-{a}/{a_\tr}}}$ in the momentum transfer rate. In the limit that the total number of degrees of freedom in the DR is large, the correction to $N_{\rm eff}$ is small. 
This model roughly captures the scaling behavior of the SPartAcous+ model~\cite{Buen-Abad:2022kgf,Buen-Abad:2023uva} in the limit of a large number of DR components.

\subsection{Impact of the Rate of Decoupling}

\begin{figure}[h!]
    \centering
    \includegraphics[width=.49\linewidth]{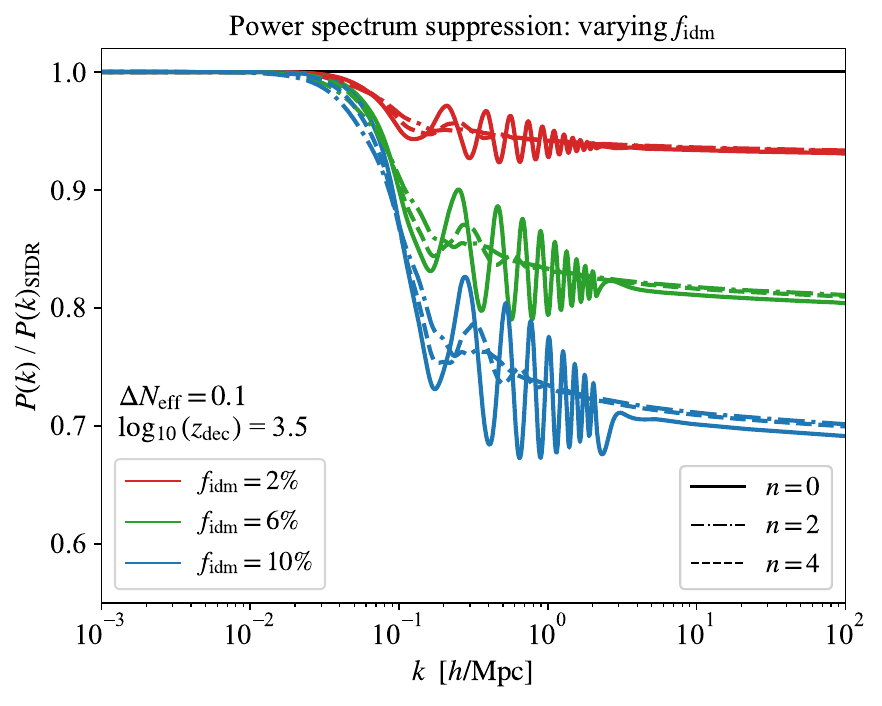}
    \includegraphics[width=.49\linewidth]{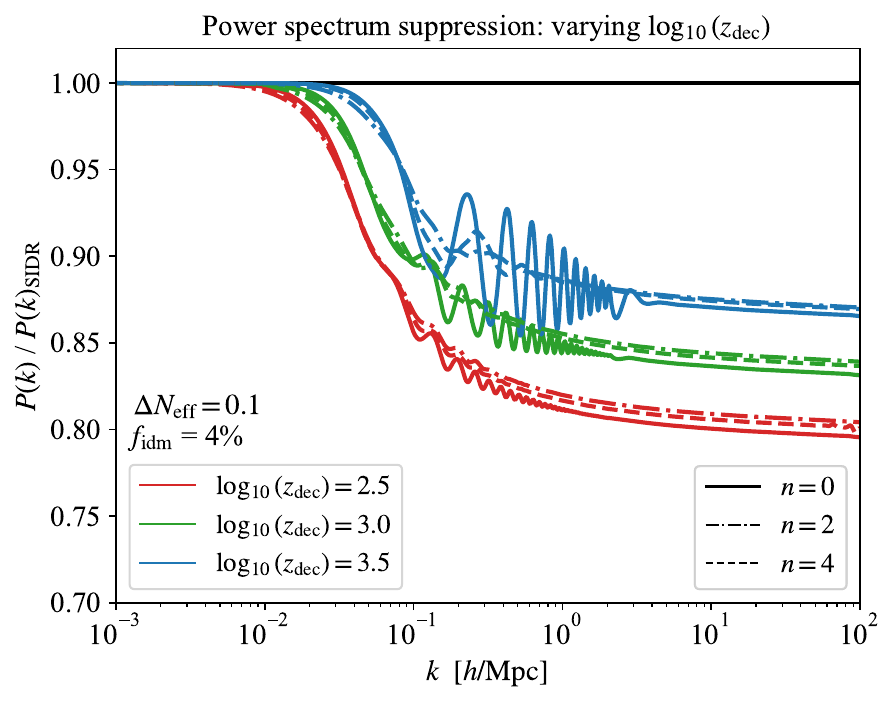}
	\caption{Ratio of the MPS of the \genidm models to that of purely self-interacting dark radiation (SIDR)~\cite{Brust:2017nmv,Blinov:2020hmc}, for different choices of $f_\idm$ (left panel) and $\logzdec$ (right panel).
    We have fixed $\DNeff = 0.1$ throughout.
    The $\theta_s$, $\omega_b$, $A_s$, $n_s$, and $\tau_{\rm reio}$ parameters have been fixed to their mean values in the Planck \lcdm fit to  TT+TE+EE+lowE+lensing+BAO data \cite{Planck:2018vyg}.
    In order to keep the time of matter-radiation equality the same as in Ref.~\cite{Planck:2018vyg} in the presence of additional DR ($z_\eq = 3387$), we choose $\omega_\dm = 0.121$.
    {\it Left}: varying $f_\idm$ from $2\%$ to $10\%$ in $4\%$ increments, while fixing $\logzdec = 3.5$.
    {\it Right}: varying $\logzdec$ from $2.5$ to $3.5$ in increments of $0.5$, while fixing $f_\idm = 4\%$.}
	\label{fig:rainbow_pk}
\end{figure}

When the interaction rate between DM and DR is much greater than Hubble, they evolve as a single tightly coupled fluid and the overall size of the interaction is irrelevant. Therefore, the primary difference between the different benchmark models is in how quickly the DM-DR interactions decouple. This affects the growth of matter perturbations, thereby affecting the MPS. Through the subsequent impact on the evolution of the perturbations in the gravitational potential, the CMB is also affected. The impact of the interactions on the MPS is noticeable even for $f_\idm \sim \mathcal{O}(1) \%$ as shown in Fig.~\ref{fig:rainbow_pk}. It is clear from the figure that the faster decoupling scenario leads to much larger amplitudes for the dark acoustic oscillations. This can be understood by recalling that these oscillations arise because, during the tightly coupled regime, the density perturbations of the iDM were oscillating due to the pressure from radiation, suppressing the growth of structure. The phase of the oscillation at decoupling is imprinted in the final power spectrum, since the regions at the maximum of the oscillation catch up with the cold DM perturbations faster after decoupling, at which point the growth factor becomes larger~\cite{Buen-Abad:2022kgf}. If the decoupling is faster, this makes the decoupling time more sharply defined, resulting in a very clear oscillatory pattern, whereas for slower decoupling there is some smearing of the effect due to the extended nature of the decoupling epoch.

\begin{figure}[h!]
    \centering
    \includegraphics[width=.49\linewidth]{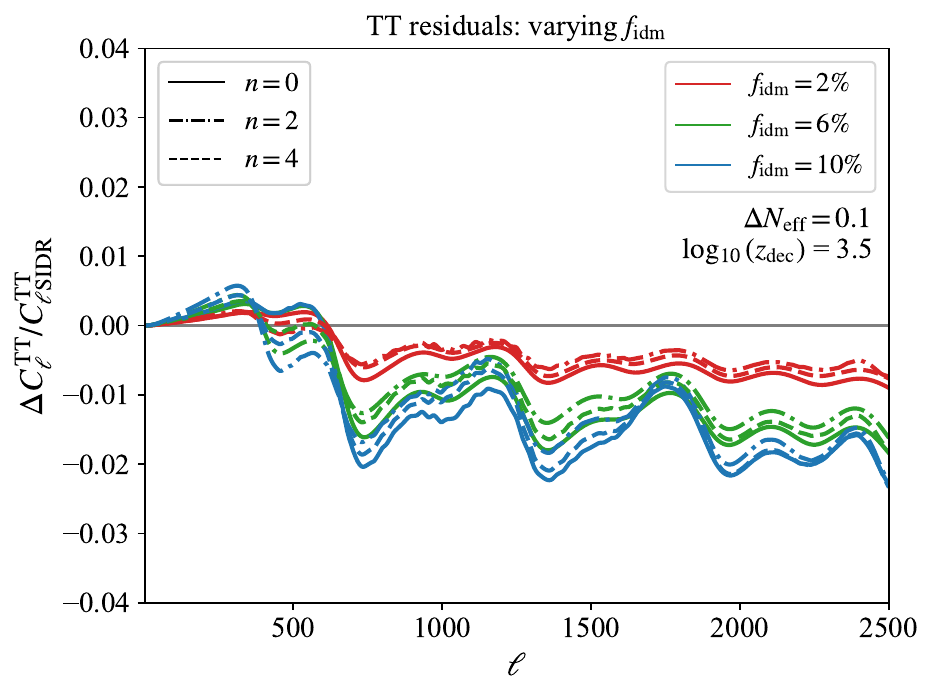}
    \includegraphics[width=.49\linewidth]{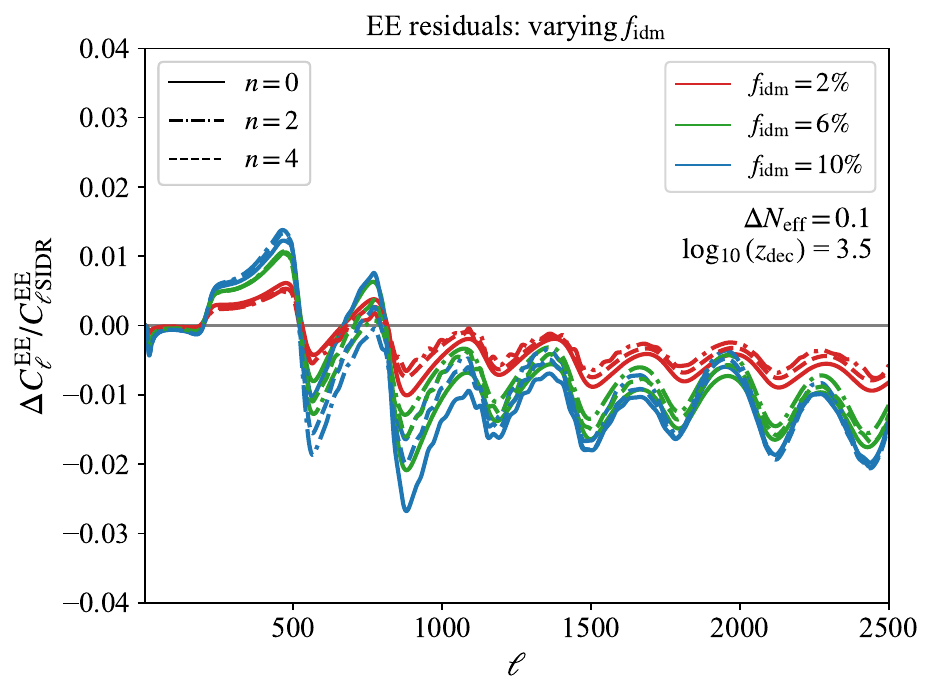}
    \includegraphics[width=.49\linewidth]{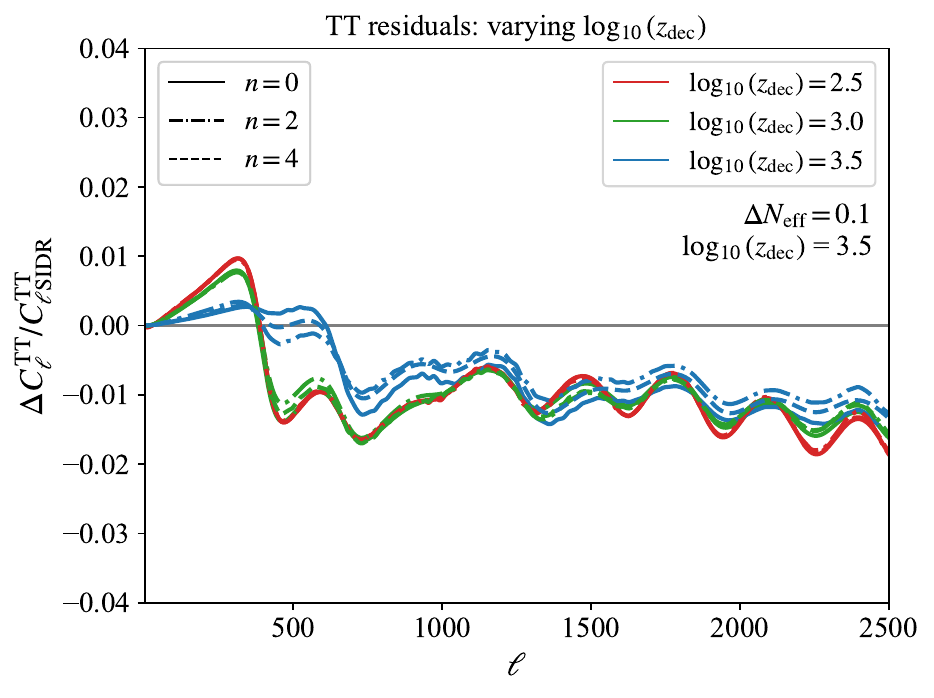}
    \includegraphics[width=.49\linewidth]{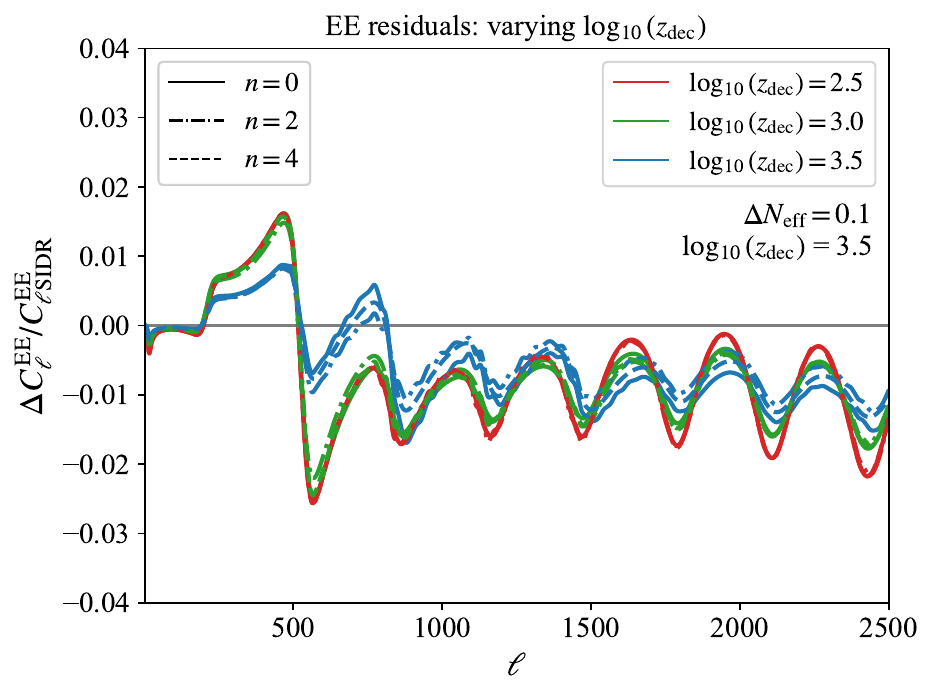}
	\caption{Residuals of the CMB $C_\ell^{TT}$ (left) and $C_\ell^{EE}$ (right) for \genidm models compared to those of SIDR, for different choices of $f_\idm$ (top) and $\logzdec$ (bottom).
    All other parameters are fixed as in Fig.~\ref{fig:rainbow_pk}.}
	\label{fig:rainbow_cmb}
\end{figure}

The impact on the CMB comes solely from the evolution and final value of the gravitational potentials~(see, e.g., Ref.~\cite{Schoneberg:2023rnx}). The rate of decoupling impacts how quickly the gravitational potentials transition to evolving as in a purely CDM scenario, which impacts the effect on the CMB power spectrum as shown in Fig.~\ref{fig:rainbow_cmb}. Qualitatively, we see that all models have a comparable impact on the CMB power spectrum, but that the amplitude of the effect is slightly larger for the $n=0$ scenario, i.e., when the decoupling is more abrupt.

\section{Methodology}
\label{sec:meth}

We devote this section to a description of our numerical analysis of the phenomenological models above.
We begin with the implementation of these models into a numerical Boltzmann solver, move on to list their parameters and the likelihood analysis tools we used to derive their posteriors, and conclude with a list of the datasets we employed in the analysis.

\subsection{Code and Parameters}

We modify the publicly available code {\tt CLASS} v3.2.5 \cite{Lesgourgues:2011re,Blas:2011rf,Lesgourgues:2011rg,Lesgourgues:2011rh}\footnote{\href{https://github.com/lesgourg/class_public}{\tt github.com/lesgourg/class\_public}.} in order to implement the $n=0,\,2$ and $4$ phenomenological models into a \genidm and solve for the cosmic evolution.
We build upon the work of the ETHOS collaboration \cite{Cyr-Racine:2015ihg}, which has already been implemented in {\tt CLASS}, to include alternative parameterizations, post-BBN thermalization of the DR, appropriate initial conditions for arbitrary $\Gamma_d$ for all $k$ modes outside the horizon, and the SPartAcous-like exponential shut-off of the $n=0$ case.
Our code is publicly available at \href{https://github.com/ManuelBuenAbad/class_gen-idm/}{\tt github.com/ManuelBuenAbad/class\_gen-idm/}.
A {\tt Mathematica} \cite{Mathematica} notebook detailing the initial conditions can be found in 
\href{https://github.com/ManuelBuenAbad/class_gen-idm/blob/main/notebooks/gen-idm_ini_perts.nb}{\tt class\_gen-idm/notebooks/gen-idm\_ini\_perts.nb} 
in our code.

In this work, we use our modified version of {\tt CLASS} together with the MCMC sampler {\tt Cobaya}~\cite{Torrado:2020dgo} to determine how the \genidm and \lcdm fits to a variety of cosmological datasets compare against each other.\footnote{
In the runs including $\mP$, we use the {\bf Halofit} model \cite{Smith:2002dz,Takahashi:2012em} to treat the MPS non-linearities. This model has been tested for \lcdm, and shown to have only a minimal impact in models where DR and iDM interact strongly (see Ref.~\cite{Schoneberg:2023rnx}), as is the case of \genidm. 
When analyzing models with $\mA$, we switch to the updated 2020 version of HMcode~\cite{Mead:2020vgs} to ensure consistency with the configuration adopted in Ref.~\cite{AtacamaCosmologyTelescope:2025nti}.}
We use the Metropolis-Hastings algorithm, taking a Gelman-Rubin (GR) convergence criterion of $R < 1.01$, where $R$ is the GR statistic~\cite{Gelman:1992zz}.
To analyze the chains and produce our figures, we make use of {\tt GetDist}~\cite{Lewis:2019xzd},

In addition to the standard \lcdm parameters $\{ \omega_b, \, \omega_\dm, \, \theta_s, \, \ln \bl( 10^{10} A_s \br), \, n_s, \, \tau_\text{reio} \}$, we scan over $f_\idm$ and $\DNeff$.
Furthermore, we fix $n$ to one of $0$, $2$, or $4$.
For $n=2$ and $n=4$ we also scan over the redshift at which the iDM decouples from the DR in log-space, namely $\log_{10} z_\dec$.
For these values of $n$, $z_{\idm,\,\dec}$ determines $\Gamma_d$ at all times.
However, since $\Gamma_d / H \approx \text{constant}$ for $n=0$, we instead scan, in log-space, over the redshift at which the exponential shutoff becomes important, namely $\log_{10} z_\tr$.
Previous work \cite{Buen-Abad:2017gxg,Buen-Abad:2022kgf,Buen-Abad:2023uva} has shown that $\Gamma_d/H \gg 1$ during RD ($z \gg z_\eq$) corresponds to a flat direction for $n=0$ models (with or without exponential shut-off), since for all such values the effective description of the iDM-DR system is that of a tightly coupled fluid.
Because of this, we simply fix $\left. \Gamma_d / H \right\vert_{\rm RD} = 10^{10}$ in our $n=0$ scans.
This value, in conjunction with $z_\tr$, determines $z_\dec$.
In summary, for $n=2$ and $n=4$ we have the additional parameters $\{ f_\idm,\, \DNeff, \log_{10} z_\dec \}$, while for $n=0$ we have $\{ f_\idm,\, \DNeff, \log_{10} z_\tr \}$.
We use the following priors in our scans:
$\log\left(10^{10} A_s\right) \in [1.61,\, 3.91]$, $n_s \in [0.8,\, 1.2]$, $\theta_s \in [0.005,\, 0.1]$, $\tau_\mathrm{reio} \in [0.01,\, 0.8]$, $\omega_b \in [0.005,\, 0.1]$, $\omega_\dm \in [0.001,\, 0.99]$, $f_\idm \in [0,\, 1]$, $\DNeff \in [0.01,\, 2]$, $\log_{10} z_\tr \in [4,\,5]$\footnote{$\log_{10} z_\tr \in [4,\,5.2]$ is chosen for $\mPF$ and $\mPHF$}, and $\log_{10}(z_\dec) \in [2,\, 3.8]$. 

\subsection{Experiments}
\label{subsec:exps}

We perform a full likelihood analysis of our \genidm model with different suites of recent cosmological datasets.
These are:

\begin{itemize}
    \item $\mP$: a mixed baseline dataset that includes the following experiments:
    \begin{itemize}
        \item {\bf Planck: } measurements of TT, TE, and EE CMB anisotropies (both high- and low-$\ell$) and lensing from Planck 2018 PR3~\cite{Planck:2018vyg}.
        \item {\bf BAO: } measurements of the BAO.
        These include measurements of $D_V/r_\mathrm{drag}$ from the Six-degree Field Galaxy Survey (6dFGS) at $z = 0.106$ \cite{Beutler:2011hx} and the Sloan Digital Sky Survey (SDSS) from the MGS galaxy sample at $z = 0.15$ \cite{Ross:2014qpa}, as well as measurements from the Data Release 16 (DR16) of the SDSS-IV extended Baryon Oscillation Spectroscopic Survey (eBOSS) \cite{eBOSS:2020yzd} of emission line galaxies (ELG) at $0.6 < z < 1.1$ \cite{eBOSS:2020qek,eBOSS:2020fvk}, quasars (QSOs) at $0.8 < z < 2.2$ \cite{eBOSS:2020gbb,eBOSS:2020uxp}, and Lyman-$\alpha$ forests \cite{eBOSS:2020tmo}.
        \item {\bf Pantheon+: } measurements of the apparent magnitudes of 1550 Type Ia supernovae (SNIa), at redshifts $0.001 < z < 2.26$ \cite{Scolnic:2021amr,Brout:2022vxf}.
    \end{itemize}
    \item $\mF$: ``Full-shape'' MPS likelihood based on BOSS DR12 LRG and eBOSS DR16 QSO data, analyzed with Effective Field Theory of Large Scale Structure (EFTofLSS) methods, and implemented in the {\tt PyBird} publicly available code~\cite{DAmico:2020kxu} ({\tt `eftboss'} and {\tt `efteboss'})\footnote{\href{https://github.com/pierrexyz/pybird}{\tt github.com/pierrexyz/pybird}.}, interfaced with {\tt Cobaya} through a custom likelihood setup.
    \item $\mH$: Late-universe measurements of the Hubble parameter $H_0$ today.
    The SH0ES collaboration obtains their value of $H_0$ from their measurements of the absolute magnitude $M_B$ of SNIa.
    Accordingly, we use a Gaussian likelihood based on their $M_B = -19.253 \pm 0.027$ result \cite{Riess:2021jrx}.
    \item $\mA$: a mixed dataset that includes:
    \begin{itemize}
        \item {\bf ACT + Planck lensing: } the Atacama Cosmology Telescope's (ACT) CMB lensing measurements from DR6 \cite{AtacamaCosmologyTelescope:2025nti,AtacamaCosmologyTelescope:2025blo}, as well as the Planck PR4 lensing likelihood \cite{Rosenberg:2022sdy,Carron:2022eyg}.
        \item {\bf Planck CMB cut used with ACT $+~\text{low-}\ell \text{ TT}$ and {\tt Sroll2} \text{EE} : } a likelihood that supplies the Planck spectra with the ACT-appropriate multipole cuts, plus low-$\ell$ TT and {\tt Sroll2} EE CMB measurements from Planck PR3. This, when combined with the ACT likelihood mentioned above, is what Refs~\cite{AtacamaCosmologyTelescope:2025nti,AtacamaCosmologyTelescope:2025blo} refer to as P-ACT.
        \item {\bf DESI: } the BAO measurements from the Dark Energy Spectroscopic Instrument (DESI) collaboration's DR2 \cite{DESI:2025zpo,DESI:2025zgx} ({\tt `bao.desi\_dr2.desi\_bao\_all'} from {\tt cobaya}'s internal likelihoods).
        \item {\bf Pantheon$+$: } the same measurements of type Ia SNe as in the $\mP$ dataset.
    \end{itemize}
\end{itemize}

Whenever we include the ACT data, we limit ourselves to the $\mA$ and $\mAH$ datasets, since due to the increased requirement on precision, the $\mA$ datasets are much more computationally expensive to analyze than $\mP$ and $\mF$.
Furthermore, we focus only on \lcdm and $n=0$ since $n=0$ is the model that fits the data best (see Tables \ref{tab:p-ph-pf-phf_val_chi2} and \ref{tab:a-ah_val_chi2}).

\section{Results}
\label{sec:results}

Let us now discuss the results from the numerical analyses described in the preceding section.
In Table~\ref{tab:p-ph-pf-phf_val_chi2} we show the $\chi^2$ values at the best fit point in parameter space for the fits of \lcdm, $n=0$, $n=2$, and $n=4$ to the $\mP$, $\mPH$, $\mPF$, and $\mPHF$ datasets.
Similarly, Table~\ref{tab:a-ah_val_chi2} shows the $\chi^2$ values at the best fit point for the \lcdm and $n=0$ model fits to the $\mA$ and $\mAH$ datasets. We can use these results to assess the preference of the various datasets for the new models over \lcdm, taking into account the three additional degrees of freedom in the \genidm models.
We do this with the help of $\Delta\chi^2$ and the Akaike Information Criterion: $\Delta\text{AIC} = \Delta\chi^2 + 2 \, \Delta \, \text{d.o.f.} = \Delta\chi^2 + 6$.
If $\Delta\text{AIC} < 0$ for a given model then the extended model is favored by the data; if on the other hand $\Delta\text{AIC} > 0$ then \lcdm is preferred.

\begin{table}[h!]
\centering
\begin{adjustbox}{max width=\columnwidth}
\begin{tabular}{@{}|c|c|c|c|ccccccc|@{}}
\toprule
\toprule
\multicolumn{1}{|c|}{Dataset} & Model & $\chi^2_\text{tot}$ & $\Delta\chi^2$ & $\chi^2_\text{CMB}$ & $\chi^2_\text{PL.lens}$ & $\chi^2_\text{BAO}$ & $\chi^2_\text{Pantheon}$ & $\chi^2_\text{SH0ES}$ & $\chi^2_{\text{EFT}_\text{BOSS}}$ & $\chi^2_{\text{EFT}_\text{eBOSS}}$ \\ \midrule
\multirow{2}{*}{$\mP$}            & $\lcdm$  & $4196.62$ & $-$ & $2762.54$ & $8.67$ & $20.86$ & $1404.55$ & $-$ & $-$ & $-$ \\
                              & $n=0$   & $4195.16$ & $-1.46$ & $2761.04$ & $8.77$ & $21.07$ & $1404.28$ & $-$ & $-$ & $-$ \\
                              & $n=2$   & $4197.20$ & $+0.58$ & $2763.58$ & $9.21$ & $19.90$ & $1404.52$ & $-$ & $-$ & $-$ \\
                              & $n=4$   & $4197.07$ & $+0.45$ & $2762.12$ & $8.92$ & $21.55$ & $1404.48$ & $-$ & $-$ & $-$ \\
                              \midrule
\multirow{2}{*}{$\mPH$}           & $\lcdm$  & $4233.91$ & $-$ & $2767.43$ & $8.69$ & $20.76$ & $1406.81$ & $30.22$ & $-$ & $-$ \\
                              & $n=0$   & $4198.93$ & $-34.98$ & $2762.96$ & $9.45$ & $20.26$ & $1405.12$ & $1.15$ & $-$ & $-$ \\
                              & $n=2$   & $4205.60$ & $-28.31$ & $2768.14$ & $9.22$ & $20.02$ & $1405.83$ & $2.39$ & $-$ & $-$ \\
                              & $n=4$   & $4202.30$ & $-31.61$ & $2765.39$ & $9.01$ & $20.24$ & $1404.67$ & $2.98$ & $-$ & $-$ \\
                              \midrule
\multirow{2}{*}{$\mPF$}           & $\lcdm$  & $4465.94$ & $-$ & $2765.63$ & $8.66$ & $20.23$ & $1405.17$ & $-$ & $175.33$ & $90.92$ \\
                              & $n=0$   & $4463.38$ & $-2.56$ & $2763.94$ & $8.81$ & $21.76$ & $1404.02$ & $-$ & $173.73$ & $91.12$ \\
                              & $n=2$   & $4465.84$ & $-0.10$ & $2765.27$ & $8.80$ & $21.61$ & $1404.27$ & $-$ & $174.56$ & $91.32$ \\
                              & $n=4$   & $4465.61$ & $-0.33$ & $2764.84$ & $8.78$ & $21.45$ & $1404.24$ & $-$ & $174.34$ & $91.96$ \\
                              \midrule
\multirow{2}{*}{$\mPHF$}         & $\lcdm$  & $4501.46$ & $-$ & $2766.32$ & $9.08$ & $20.48$ & $1406.07$ & $33.64$ & $175.06$ & $90.81$ \\
                              & $n=0$   & $4474.32$ & $-27.14$ & $2767.31$ & $9.27$ & $19.85$ & $1405.93$ & $4.77$ & $175.13$ & $92.07$ \\
                              & $n=2$   & $4476.50$ & $-24.96$ & $2772.20$ & $9.21$ & $20.32$ & $1405.64$ & $2.54$ & $175.33$ & $91.26$ \\
                              & $n=4$   & $4475.79$ & $-25.67$ & $2770.93$ & $8.22$ & $20.45$ & $1406.32$ & $2.20$ & $175.38$ & $92.29$ \\ 
                              \bottomrule \bottomrule
\end{tabular}
\end{adjustbox}
\caption{Table of the best-fit $\chi^2$ values of the $n=0,2,4$ and $\lcdm$ models for the datasets studied in this work.}
\label{tab:p-ph-pf-phf_val_chi2}
\end{table}
%

%

%
\begin{table}[h!]
\centering
\begin{adjustbox}{max width=\columnwidth}
\begin{tabular}{@{}|c|c|c|c|ccccccc|@{}}
\toprule
\toprule
\multicolumn{1}{|c|}{Dataset} & Model & $\chi^2_\text{tot}$ & $\Delta\chi^2$ & $\chi^2_\text{ACT}$ & $\chi^2_\text{PL.cut}$ & $\chi^2_\text{ACT.lens}$ & $\chi^2_\text{PL.lens}$ & $\chi^2_\text{DESI}$ & $\chi^2_\text{Pantheon}$ & $\chi^2_\text{SH0ES}$ \\ \midrule
\multirow{2}{*}{$\mA$}         & $\lcdm$  & $7964.76$ & $-$ & $5888.99$ & $635.14$ & $14.12$ & $8.31$ & $12.44$ & $1405.76$ & $-$ \\
                              & $n=0$   & $7964.39$ & $-0.37$ & $5888.48$ & $635.50$ & $14.10$ & $8.31$ & $12.15$ & $1405.85$ & $-$ \\
                              \midrule
\multirow{2}{*}{$\mAH$}         & $\lcdm$  & $7996.40$ & $-$ & $5889.28$ & $636.68$ & $14.38$ & $8.32$ & $10.57$ & $1407.20$ & $29.97$ \\
                              & $n=0$   & $7984.08$ & $-12.32$ & $5894.06$ & $636.75$ & $14.88$ & $8.40$ & $10.58$ & $1406.87$ & $11.94$\\
                              \bottomrule \bottomrule
\end{tabular}
\end{adjustbox}
\caption{Table of the best-fit $\chi^2$ values of the $n=0$ and $\lcdm$ models for the datasets studied in this work.}
\label{tab:a-ah_val_chi2}
\end{table}

\begin{table}[h!]
\centering
\begin{adjustbox}{max width=\columnwidth}
\begin{tabular}{|c|c|c|c|c|c|c|}
\toprule
$\mP$        & $\Delta\chi^2$ &  $\DNeff$ & $f_\idm [\%]$ & \logzdec & $H_0~[\km/\seg/\Mpc]$ &  $S_8$ \\ \midrule
\lcdm   & $-$ & $-$ & $-$ & $-$ & $67.7\, (67.6^{+0.8}_{-0.8})$ & $0.825\,(0.827^{+0.020}_{-0.019})$ \\
$n=0$   & $-1.46$ & $0.38\,(0.35^{+0.40}_{-0.34})$ & $2.7\,(1.7^{+2.1}_{-1.7})$ & $3.3\,(3.3^{+0.3}_{-0.5})$ & $69.8\,(69.5^{+2.6}_{-2.3})$ & $0.822\,(0.824^{+0.020}_{-0.021})$\\
$n=2$   & $+0.58$ & $0.17\,(0.24^{+0.29}_{-0.23})$ & $0.4\,(1.1^{+2.1}_{-1.1})$ & $3.3\,(3.1^{+0.7}_{-1.1})$ & $68.7\,(69.0^{+2.1}_{-1.8})$ & $0.833\,(0.823^{+0.022}_{-0.023})$ \\
$n=4$   & $+0.45$ & $0.18\,(0.25^{+0.31}_{-0.24})$ & $1.0\,(1.1^{+2.0}_{-1.1})$ & $3.3\,(3.1^{+0.7}_{-0.9})$ & $68.5\,(69.1^{+2.1}_{-1.8})$ & $0.829\,(0.823^{+0.020}_{-0.022})$\\
\bottomrule
\end{tabular}
\end{adjustbox}
\caption{Results of the $n=0,2,4$ and \lcdm fits to the $\mP$ dataset.
Shown are the best fit $\Delta\chi^2$ values relative to \lcdm, as well as the best fit values (mean $\pm 95\%$ confidence regions) for the $\DNeff$, $f_\idm$, \logzdec, $H_0$, and $S_8$ parameters.
To obtain the best fit AIC number, simply add $+6$ to the $\Delta\chi^2$ values shown here.
}
\label{tab:P_95}
\end{table}

\begin{table}[h!]
\centering
\begin{adjustbox}{max width=\columnwidth}
\begin{tabular}{|c|c|c|c|c|c|c|}
\toprule
$\mPH$        & $\Delta\chi^2$ &  $\DNeff$ & $f_\idm [\%]$ & \logzdec & $H_0~[\km/\seg/\Mpc]$ &  $S_8$ \\ \midrule
\lcdm   & $-$ & $-$ & $-$ & $-$ & $68.6\,(68.5^{+0.7}_{-0.7})$ & $0.810\,(0.810^{+0.018}_{-0.018})$ \\
$n=0$   & $-34.98$ & $0.89\,(0.86^{+0.31}_{-0.30})$ & $2.9\,( 3.3^{+2.2}_{-2.2})$ & $3.3\,(3.3^{+0.1}_{-0.1})$ & $72.6\,(72.6^{+1.6}_{-1.6})$ & $0.820\,(0.820^{+0.019}_{-0.019})$ \\
$n=2$   & $-28.31$ & $0.73\,(0.70^{+0.29}_{-0.27})$ & $1.8\,(1.8^{+2.4}_{-1.8})$ & $3.2\,(3.1^{+0.7}_{-0.8})$ & $72.2\,(72.0^{+1.6}_{-1.5})$ & $0.817\,(0.817^{+0.021}_{-0.022})$ \\
$n=4$   & $-31.61$ & $0.79\,(0.70^{+0.29}_{-0.27})$ & $2.7\,(1.8^{+2.4}_{-1.8})$ & $3.3\,(3.1^{+0.7}_{-0.8})$ & $72.0\,(72.0^{+1.6}_{-1.5})$ & $0.824\,(0.817^{+0.021}_{-0.022})$ \\
\bottomrule
\end{tabular}
\end{adjustbox}
\caption{Results of the $n=0,2,4$ and \lcdm fits to the $\mPH$ dataset.
Shown are the best fit $\Delta\chi^2$ values relative to \lcdm, as well as the best fit values (mean $\pm 95\%$ confidence regions) for the $\DNeff$, $f_\idm$, \logzdec, $H_0$, and $S_8$ parameters.
To obtain the best fit AIC number, simply add $+6$ to the $\Delta\chi^2$ values shown here.
}
\label{tab:PH_95}
\end{table}

\begin{table}[h!]
\centering
\begin{adjustbox}{max width=\columnwidth}
\begin{tabular}{|c|c|c|c|c|c|c|}
\toprule
$\mPF$        & $\Delta\chi^2$ &  $\DNeff$ & $f_\idm [\%]$ & \logzdec & $H_0~[\km/\seg/\Mpc]$ &  $S_8$ \\ \midrule
\lcdm   & $-$ & $-$ & $-$ & $-$ & $68.1\,(67.8^{+0.7}_{-0.7})$ & $0.811\,(0.823^{+0.018}_{-0.018})$ \\
$n=0$   & $-2.56$ & $0.06\,(0.25^{+0.31}_{-0.24})$ & $1.1\,(0.9^{+1.3}_{-0.9})$ & $3.6\,(3.4^{+0.5}_{-0.7})$ & $67.9\,(69.2^{+2.3}_{-1.8})$ & $0.823\,(0.820^{+0.019}_{-0.020})$ \\
$n=2$   & $-0.10$ & $ 0.02\,(0.22^{+0.26}_{-0.21})$ & $2.0\,(1.4^{+2.5}_{-1.4})$ & $3.5\,(3.3^{+0.5}_{-1.0})$ & $68.4\,(69.1^{+1.9}_{-1.6})$ & $0.817\,(0.817^{+0.021}_{-0.023})$ \\
$n=4$   & $-0.33$ & $0.06\,(0.22^{+0.27}_{-0.21})$ & $0.1\,(1.2^{+1.9}_{-1.2})$ & $3.6\,(3.2^{+0.6}_{-1.0})$ & $68.0\,(69.0^{+2.0}_{-1.7})$ & $0.819\,(0.818^{+0.021}_{-0.022})$ \\
\bottomrule
\end{tabular}
\end{adjustbox}
\caption{Results of the $n=0,2,4$ and \lcdm fits to the $\mPF$ dataset.
Shown are the best fit $\Delta\chi^2$ values relative to \lcdm, as well as the best fit values (mean $\pm 95\%$ confidence regions) for the $\DNeff$, $f_\idm$, \logzdec, $H_0$, and $S_8$ parameters.
To obtain the best fit AIC number, simply add $+6$ to the $\Delta\chi^2$ values shown here.
}
\label{tab:PF_95}
\end{table}

\begin{table}[h!]
\centering
\begin{adjustbox}{max width=\columnwidth}
\begin{tabular}{|c|c|c|c|c|c|c|}
\toprule
$\mPHF$        & $\Delta\chi^2$ &  $\DNeff$ & $f_\idm [\%]$ & \logzdec & $H_0~[\km/\seg/\Mpc]$ &  $S_8$ \\ \midrule
\lcdm   & $-$ & $-$ & $-$ & $-$ & $68.4\,(68.5^{+0.7}_{-0.6})$ & $0.811\,(0.809^{+0.018}_{-0.017})$ \\
$n=0$   & $-27.14$ & $0.62\,(0.66^{+0.27}_{-0.27})$ & $0.8\,(1.0^{+1.0}_{-1.0})$ & $3.3\,(3.2^{+0.6}_{-0.5})$ & $71.6\,(71.9^{+1.5}_{-1.5})$ & $0.820\,(0.818^{+0.018}_{-0.018})$\\
$n=2$   & $-24.96$ & $0.73\,(0.66^{+0.27}_{-0.27})$ & $2.9\,(1.9^{+2.3}_{-1.9})$ & $3.4\,(3.2^{+0.6}_{-0.8})$ & $72.1\,(71.9^{+1.5}_{-1.5})$ & $0.813\,(0.813^{+0.019}_{-0.021})$ \\
$n=4$   & $-25.67$ & $0.71\,(0.65^{+0.26}_{-0.26})$ & $1.2\,(1.3^{+1.4}_{-1.3})$ & $3.2\,(3.1^{+0.7}_{-0.8})$ & $72.2\,(71.8^{+1.4}_{-1.5})$ & $0.812\,(0.817^{+0.018}_{-0.019})$ \\
\bottomrule
\end{tabular}
\end{adjustbox}
\caption{Results of the $n=0,2,4$ and \lcdm fits to the $\mPHF$ dataset.
Shown are the best fit $\Delta\chi^2$ values relative to \lcdm, as well as the best fit values (mean $\pm 95\%$ confidence regions) for the $\DNeff$, $f_\idm$, \logzdec, $H_0$, and $S_8$ parameters.
To obtain the best fit AIC number, simply add $+6$ to the $\Delta\chi^2$ values shown here.
}
\label{tab:PHF_95}
\end{table}

\begin{table}[h!]
\centering
\begin{adjustbox}{max width=\columnwidth}
\begin{tabular}{|c|c|c|c|c|c|c|}
\toprule
Datasets        & $\Delta\chi^2$ &  $\DNeff$ & $f_\idm [\%]$ & \logzdec & $H_0~[\km/\seg/\Mpc]$ &  $S_8$ \\ \midrule
\lcdm ($\mA$)   & $-$ &  $-$ & $-$ & $-$ & $68.4\,(68.3^{+0.5}_{-0.5})$ & $0.814\,(0.815^{+0.014}_{-0.013})$ \\
$n=0$~~~($\mA$)   & $-0.37$ & $0.02\,(0.10^{+0.16}_{-0.09})$ & $0.0\,(0.9^{+1.5}_{-0.9})$ & $3.4\,(3.3^{+0.4}_{-0.5})$ & $68.5\,(68.9^{+1.2}_{-0.9})$ & $0.815\,(0.812^{+0.015}_{-0.016})$ \\ \midrule
~\lcdm ($\mAH$)   & $-$ & $-$ & $-$ & $-$ & $68.7\,(68.7^{+0.5}_{-0.5})$ & $0.808\,(0.807^{+0.013}_{-0.013})$ \\
~$n=0$~~~($\mAH$)   & $-12.36$ & $0.31\,(0.38^{+0.33}_{-0.27})$ & $1.3\,(1.5^{+1.9}_{-1.5})$ & $3.4\,(3.3^{+0.3}_{-0.4})$ & $70.4\,(70.7^{+1.8}_{-1.4})$ & $0.809\,(0.809^{+0.014}_{-0.014})$ \\
\bottomrule
\end{tabular}
\end{adjustbox}
\caption{Results of the $n=0$ and \lcdm fits to the $\mA$ and $\mAH$ datasets.
Shown are the best fit $\Delta\chi^2$ values relative to \lcdm, as well as the best fit values (mean $\pm 95\%$ confidence regions) for the $\DNeff$, $f_\idm$, \logzdec, $H_0$, and $S_8$ parameters.
To obtain the best fit AIC number, simply add $+6$ to the $\Delta\chi^2$ values shown here.
}
\label{tab:A_95}
\end{table}

From the results in Tables~\ref{tab:p-ph-pf-phf_val_chi2} and~\ref{tab:a-ah_val_chi2}, it is clear that for the datasets without the local SH0ES measurement (\eg, $\mP$, and $\mPF$), the penalty from having additional degrees of freedom outweighs the negligible changes to $\Delta\chi^2$, which means that \lcdm remains the preferred model. 
When including the SH0ES data (in $\mPH$, and $\mPHF$), which is famously in tension within \lcdm, all three $n$ variants yield negative values of $\Delta\text{AIC}$, with a clear preference for the $n=0$ model ($\Delta\text{AIC} \approx -29$ for $\mPH$, $\Delta\text{AIC} \approx -21$ for $\mPHF$). Although the improvements in the fit are dominated by a better fit to the SH0ES data, it is noteworthy that for the $n=0$ fit to $\mPH$ the $\chi^2$ from the fit to the CMB data is effectively the same as for the \lcdm fit to just $\mP$, showing that the model can accommodate the larger value of $H_0$ without sacrificing the goodness of the fit to the CMB. The resulting negative values of $\Delta\text{AIC}$ therefore reflect the combined datasets and should not be interpreted as evidence that the CMB
data alone strongly prefer the interacting model over \lcdm. For the $\mA$ and $\mAH$ datasets we observe a similar trend in results: the AIC indicates $\mA$ prefers the \lcdm model, whereas $\mAH$ prefers \genidm. However, in this case the improvement is more modest and comes solely from the improvement in fitting SH0ES data, at the expense of a slight worsening of the fit to the CMB.
We can then deduce that all the $n$ variants of our \genidm model are preferred only when the local $H_0$ measurements are included, and that otherwise \lcdm is preferred.

Tables~\ref{tab:P_95}, \ref{tab:PH_95}, \ref{tab:PF_95} and \ref{tab:PHF_95} show $\Delta\chi^2$ once again, as well as the best fit, mean, and 95\% confidence interval values for the parameters of the models considered: $\DNeff$, $f_\idm$, and $\log_{10}\bl( z_\dec \br)$, as well as $H_0$ and $S_8$, for the datasets $\mP$, $\mPH$, $\mPF$ and $\mPHF$ for all models studied~(see Appendix~\ref{appendix} for results including all the parameters of each model). Table~\ref{tab:A_95} shows the same parameters for the fits to $\mA$ and $\mAH$ for \lcdm and for the $n=0$ model.
As expected, the inclusion of $\mH$, which is associated with a strong preference for \genidm over \lcdm, results in mean values for $\DNeff$ significantly above $0$.
The extra amount of radiation decreases the size of the sound horizon at the time of recombination, which is correlated with an increase in $H_0$ \cite{Aylor:2018drw,Knox:2019rjx}. On the other hand, the values of $f_\idm$ remain consistent with $0$ at $2\sigma$ for all models and datasets, except the fit of $n=0$ to $\mPH$. Note that the $n=0$ case allows fairly large values of $\DNeff$ even without the inclusion of the SH0ES data, showing that abrupt decoupling of the iDM component from DR substantially weakens the constraints on DR.

\begin{figure}[h!]
    \centering
    \includegraphics[width=.75\linewidth]{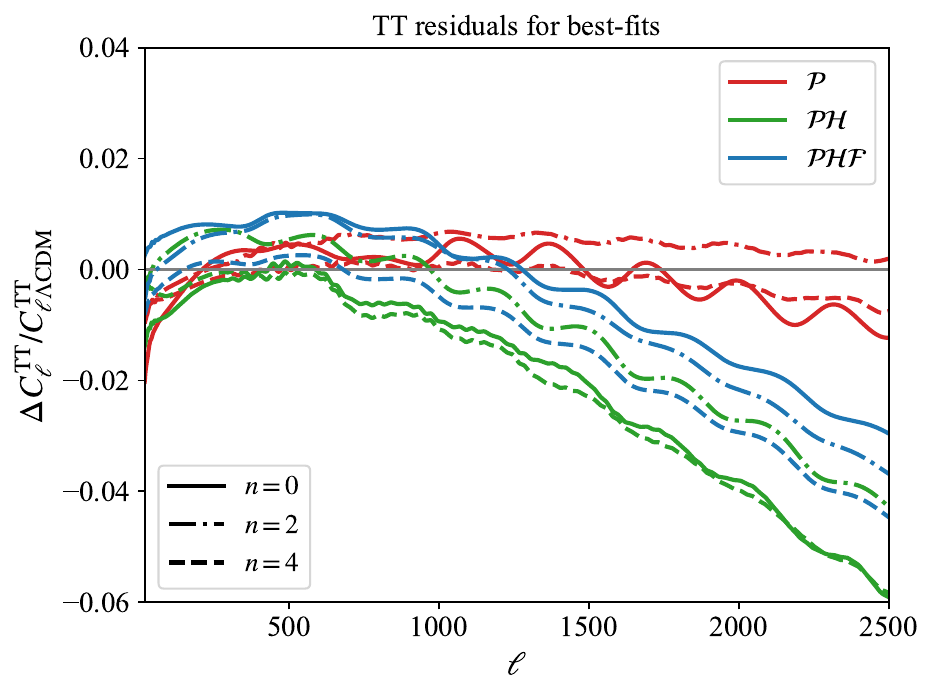}
    \includegraphics[width=.75\linewidth]{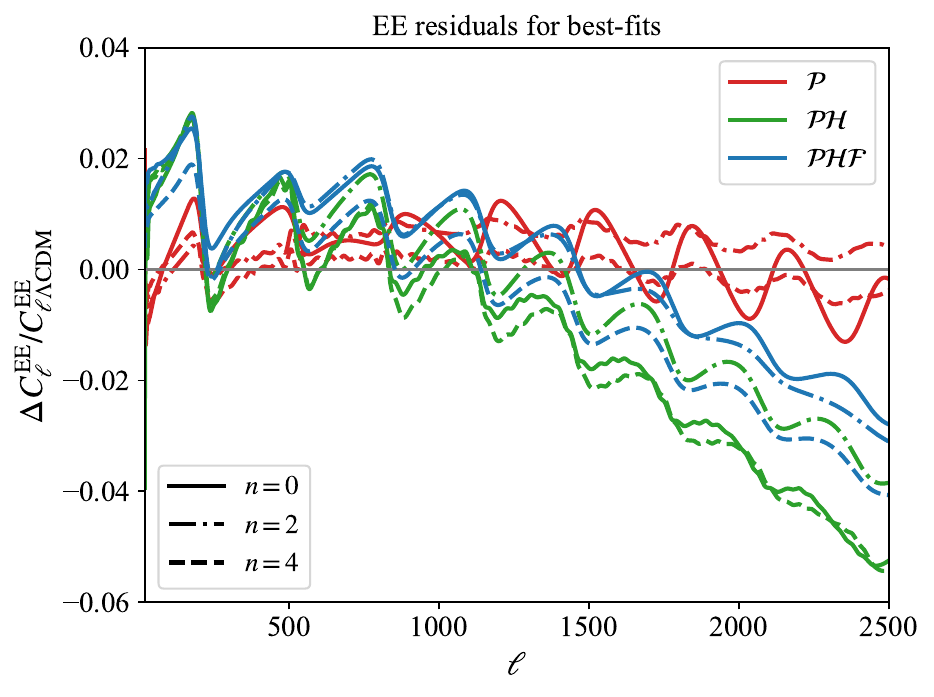}
	\caption{$C_\ell^{TT}$ (top) and $C_\ell^{EE}$ (bottom) residuals of the \genidm models compared to \lcdm, using the best fit point of each model to the $\mP$ (red), $\mPH$ (green), and $\mPHF$ (blue) datasets.}
	\label{fig:bf_cmb1}
\end{figure}

\begin{figure}[h!]
    \centering
    \includegraphics[width=.49\linewidth]{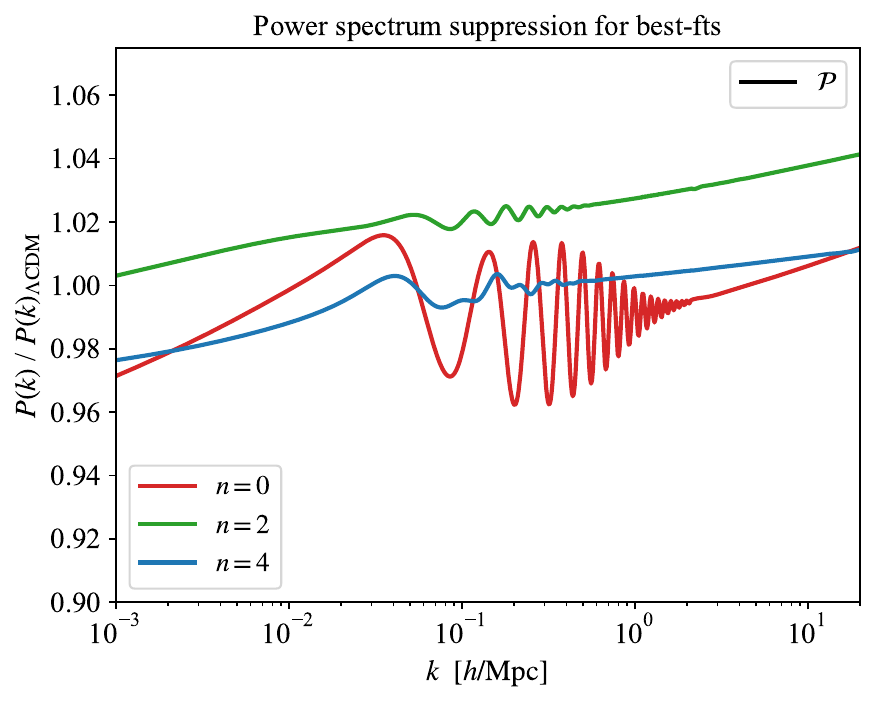}
    \includegraphics[width=.49\linewidth]{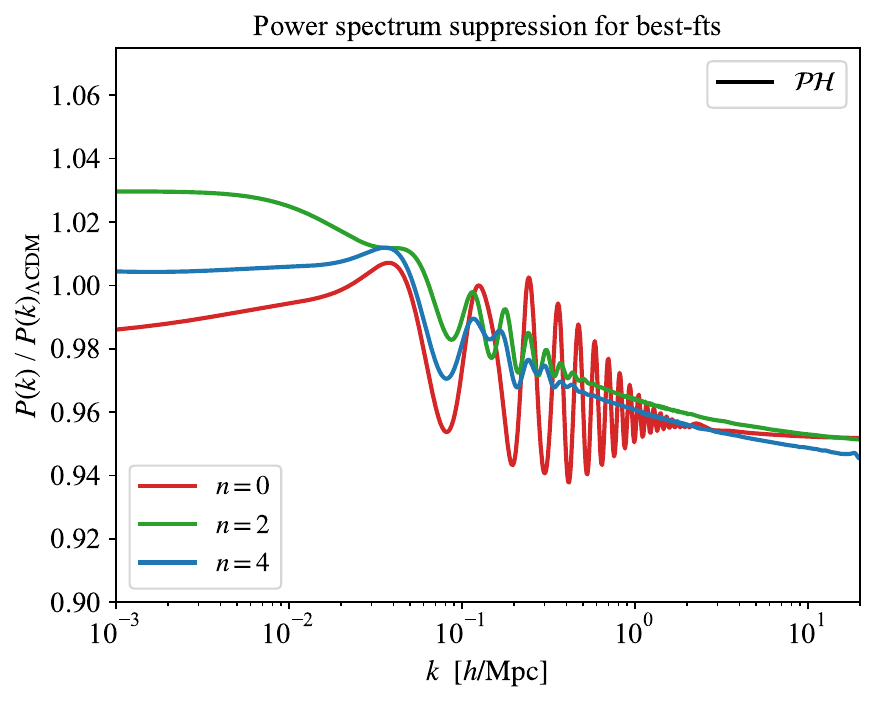}
    \includegraphics[width=.49\linewidth]{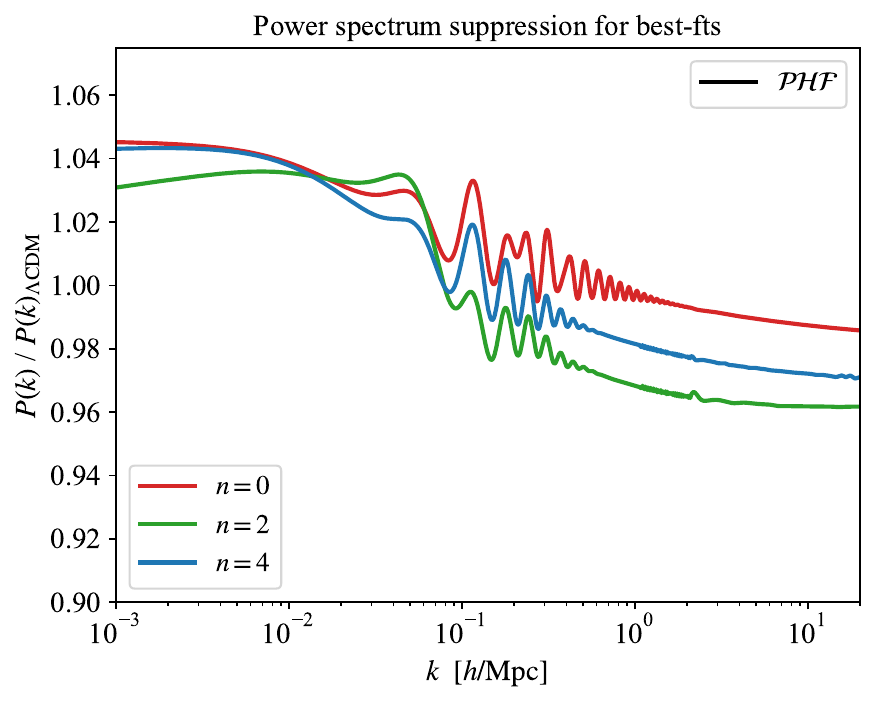}
	\caption{Ratios of the linear MPS in the \genidm models to that in the \lcdm model, for the best fit point of each model to the $\mP$ (upper left), $\mPH$ (upper right), and $\mPHF$ (lower) datasets. In all curves one can clearly see the effect of the dark acoustic oscillations. In some of the figures, one can also observe a little interference between the dark acoustic oscillations and BAO.}
	\label{fig:bf_pk1}
\end{figure}

In Fig.~\ref{fig:bf_cmb1}, we can see the residuals of $C_\ell^{TT}$ and $C_\ell^{EE}$ for the best fits of the \genidm variants to each dataset compared to the best fit of \lcdm to the same dataset. It is worth noting that for the fits to $\mP$, the power spectrum shows no noticeable decrease at high $\ell$ despite the expected enhanced Silk damping from non-zero \DNeff. This indicates that over the scales probed by Planck, the scale dependent effects due to the iDM-iDR interactions allows the best fit of the model to avoid the decrease in power expected from extra contributions to $\DNeff$. For the fits that include the SH0ES prior, \DNeff is pushed to higher values, and the figure shows that this indeed leads to a suppression at high $\ell$ as expected from the increased damping. Despite this increased suppression, the CMB $\chi^2$ values in
Table~\ref{tab:p-ph-pf-phf_val_chi2} show that, especially for the $n=0$
model, this does not lead to a significant worsening of the fit to the CMB. In Fig.~\ref{fig:bf_pk1}, we see the effects of each model on the MPS when fit to $\mP$, $\mPH$ and $\mPHF$ data combinations. In all cases, one can see the effects of the interactions in the form of dark acoustic oscillations, which are more pronounced in the $n=0$ case since the fast decoupling leads to less damping of the acoustic feature. In the fits to $\mPH$ and $\mPHF$ one can also note the suppression at smaller scales due to the interactions, while in $\mP$ there is a small enhancement despite the effect of the interactions, due to the impact of increased $n_s$. 

\begin{figure}[h!]
    \centering
    \includegraphics[width=.75\linewidth]{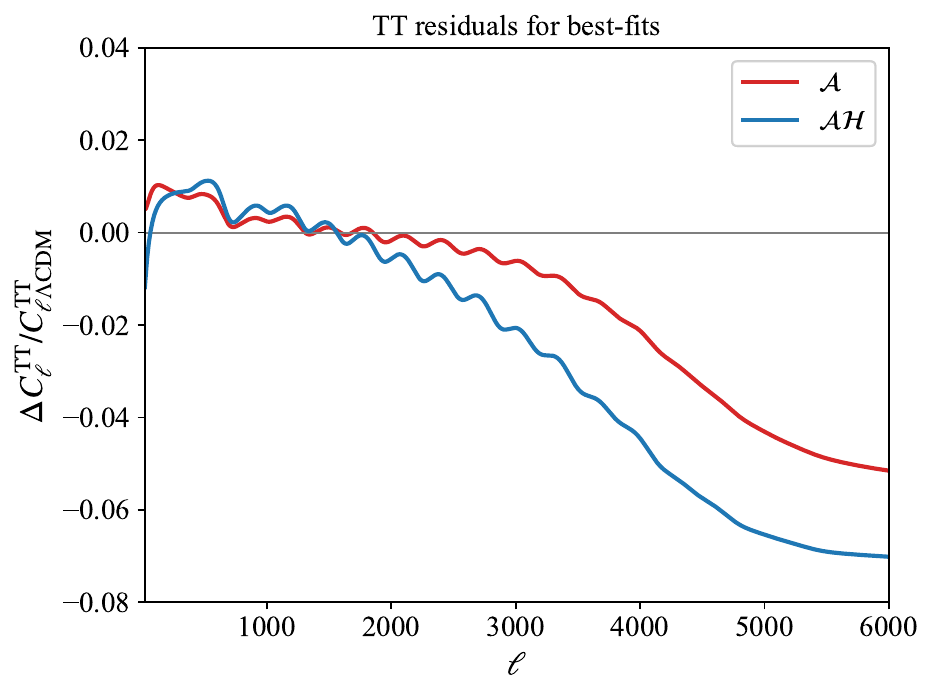}
    \includegraphics[width=.75\linewidth]{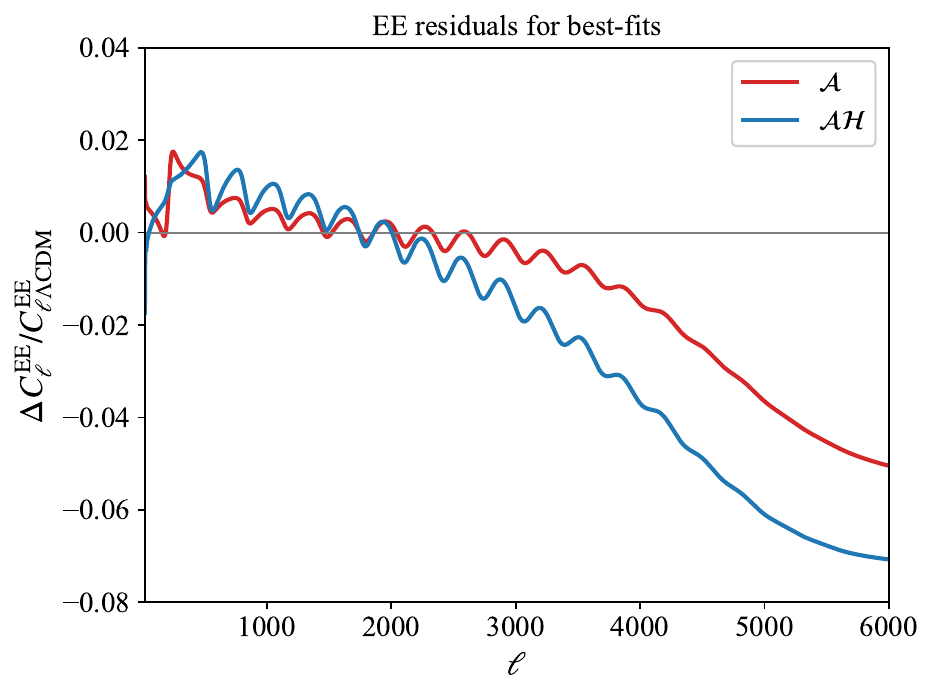}
	\caption{$C_\ell^{TT}$ (top) and $C_\ell^{EE}$ (bottom) residuals of the $n=0$ model compared to \lcdm, for the best fit point to the $\mA$ (red) and $\mAH$ (blue) datasets.}
	\label{fig:bf_cmb2}
\end{figure}

In Fig.~\ref{fig:bf_cmb2}, we show the residuals between \lcdm and the $n=0$ model for data combinations that include the new ACT results. Since ACT extends the sensitivity to high $\ell$, this probes the regions in which higher $\DNeff$ causes increased Silk damping, as had already been seen in Fig.~\ref{fig:bf_cmb1}. Comparing both figures, we see that the ACT results do not favor a decrease of power at small scales, leading to tighter constraints on $\DNeff$. This is in agreement with what had been observed for simple models of DR in Ref.~\cite{AtacamaCosmologyTelescope:2025nti}. The inclusion of the SH0ES results pushes the value of $\DNeff$ higher, leading to increased suppression of the high $\ell$ power spectrum, as seen in the figure, but since this is not preferred by the ACT data, it cannot accommodate values of $\DNeff$ as large as what is found when using only Planck data.

\begin{figure}[h!]
    \centering
    \includegraphics[width=.75\linewidth]{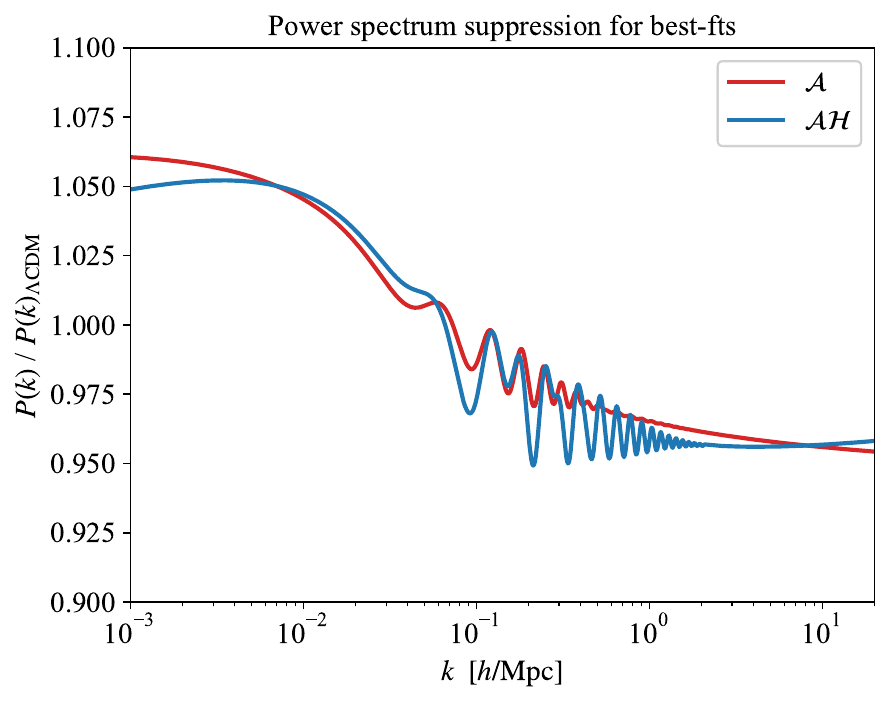}
	\caption{Ratios of the linear MPS in the $n=0$ model to that in the \lcdm model, for the best fit point to the $\mA$ (red) and $\mAH$ (blue) datasets.}
	\label{fig:bf_pk2}
\end{figure}

\begin{figure}[tbh!]
    \centering
	\includegraphics[width=0.75\linewidth]{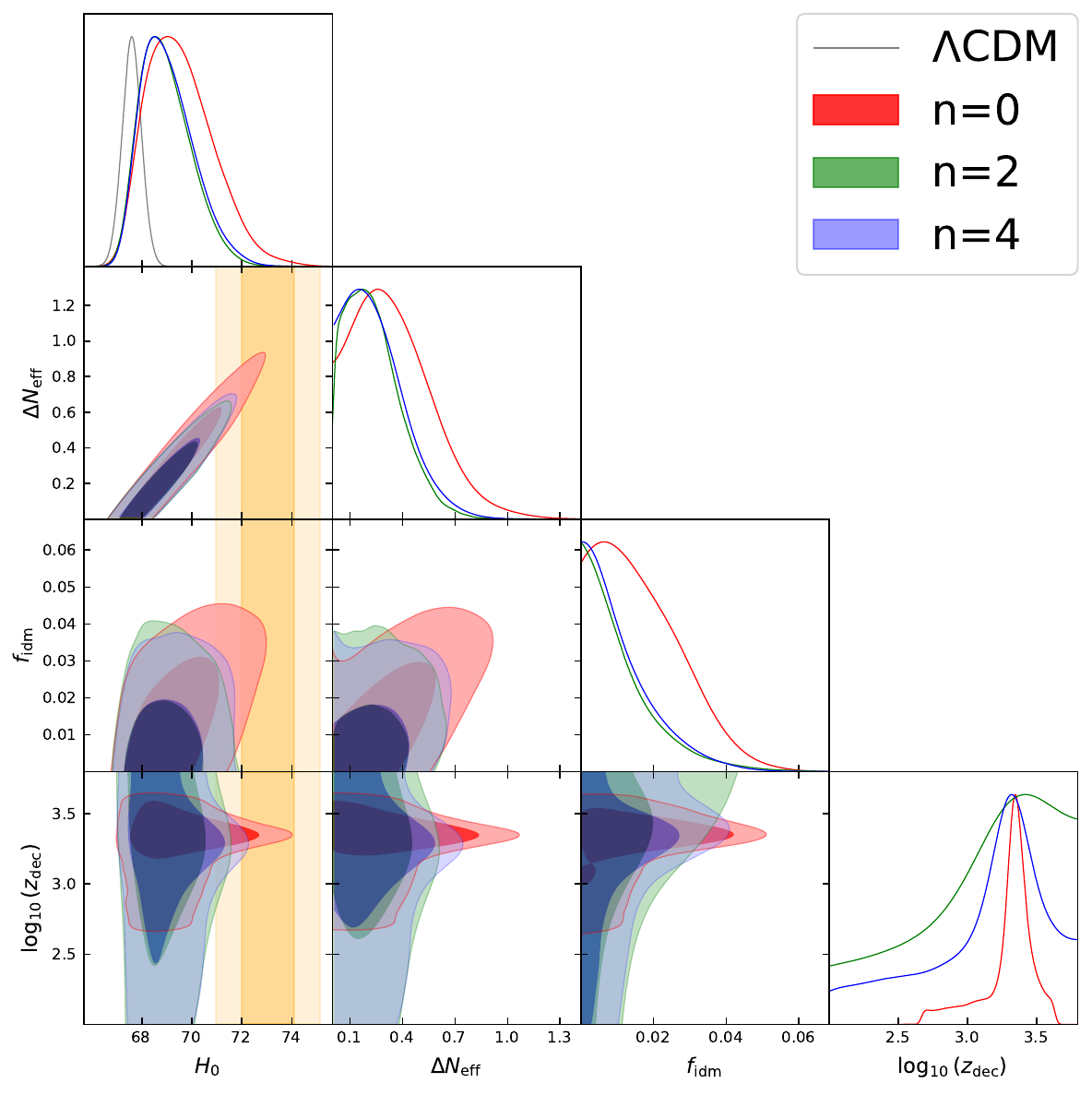}
	\caption{
    1D and 2D posterior distributions of the $H_0$, $\DNeff$, $f_\idm$, and $\log_{10} (z_{\dec})$ parameters of the \lcdm (gray), $n=0$ (red), $n=2$ (green), and $n=4$ (blue) models, fitted to the $\mP$ dataset. The orange bands show the $1\sigma$ and $2\sigma$ contours of the $H_0$ measurement by the SH0ES collaboration \cite{Riess:2021jrx}.
    }
	\label{fig:P-triangles}
\end{figure}

\begin{figure}[tbh!]
    \centering
	\includegraphics[width=0.75\linewidth]{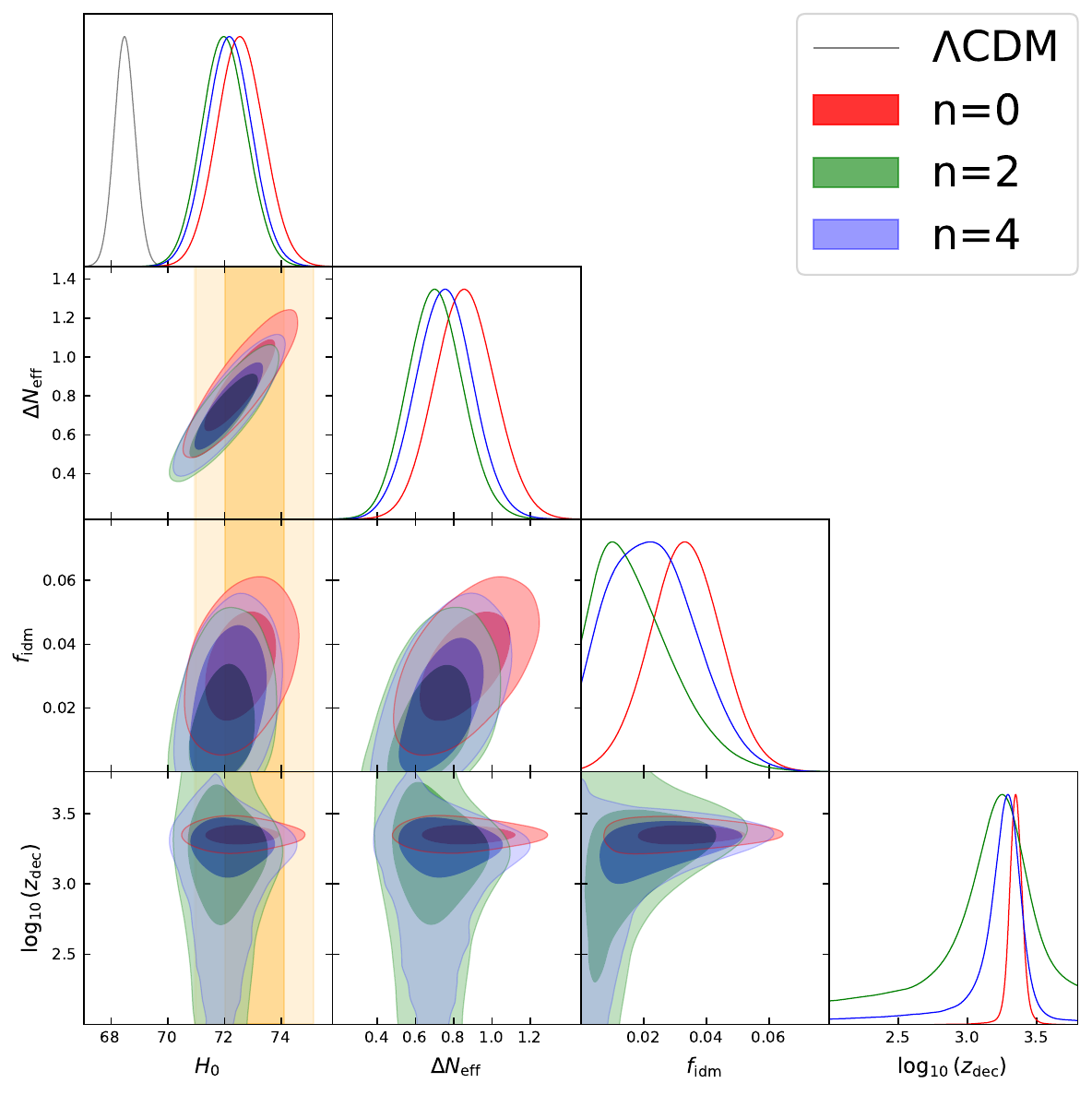}
	\caption{
    1D and 2D posterior distributions of the $H_0$, $\DNeff$, $f_\idm$, and $\log_{10} (z_{\dec})$ parameters of the \lcdm (gray), $n=0$ (red), $n=2$ (green), and $n=4$ (blue) models, fitted to the $\mPH$ dataset. The orange bands show the $1\sigma$ and $2\sigma$ contours of the $H_0$ measurement by the SH0ES collaboration \cite{Riess:2021jrx}.
    }
	\label{fig:PH-triangles}
\end{figure}

\begin{figure}[tbh!]
    \centering
	\includegraphics[width=0.75\linewidth]{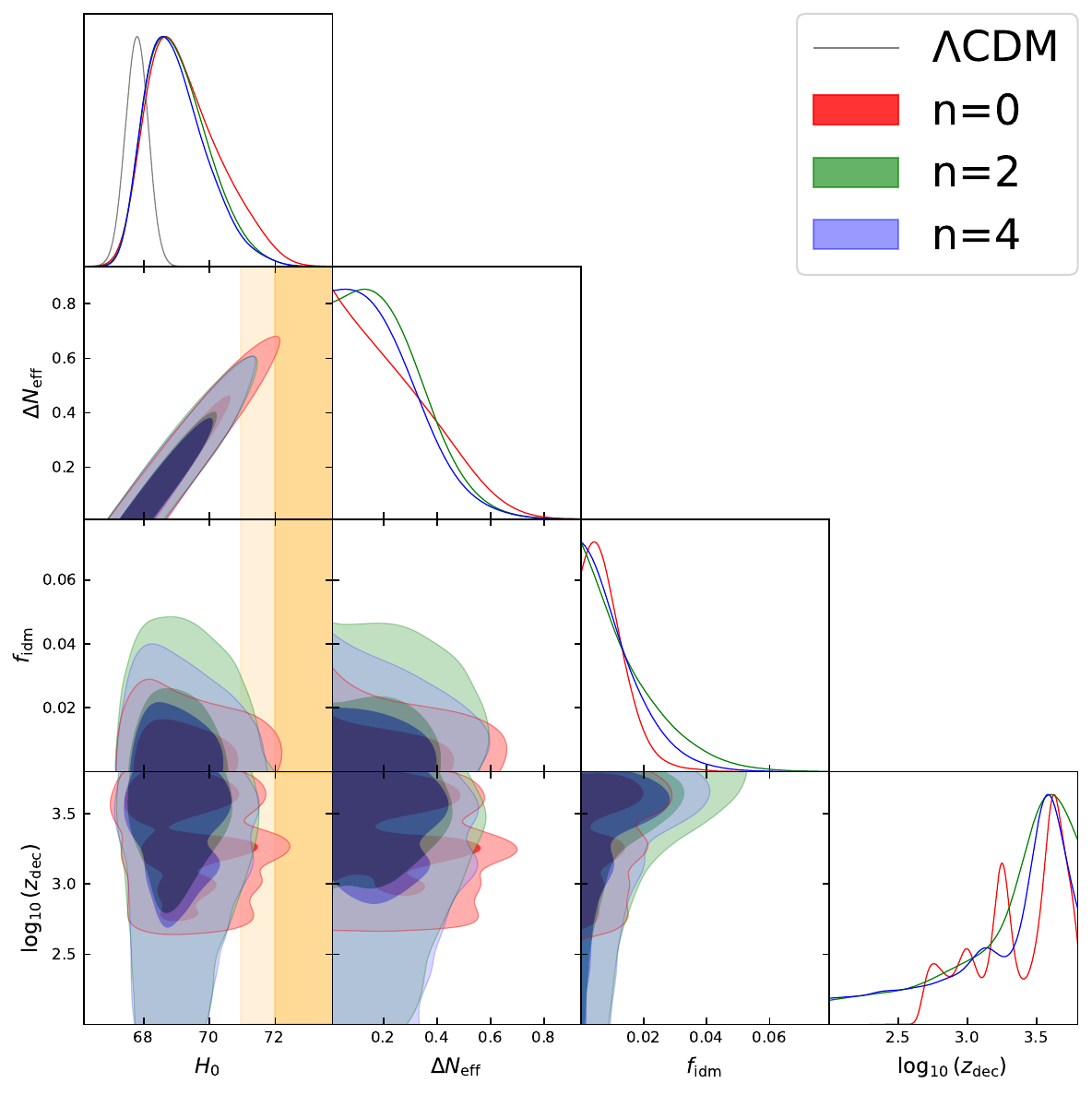}
	\caption{
    1D and 2D posterior distributions of the $H_0$, $\DNeff$, $f_\idm$, and $\log_{10} (z_{\dec})$ parameters of the \lcdm (gray), $n=0$ (red), $n=2$ (green), and $n=4$ (blue) models, fitted to the $\mPF$ dataset. The orange bands show the $1\sigma$ and $2\sigma$ contours of the $H_0$ measurement by the SH0ES collaboration \cite{Riess:2021jrx}.
    }
	\label{fig:PF-triangles}
\end{figure}

\begin{figure}[tbh!]
    \centering
	\includegraphics[width=0.75\linewidth]{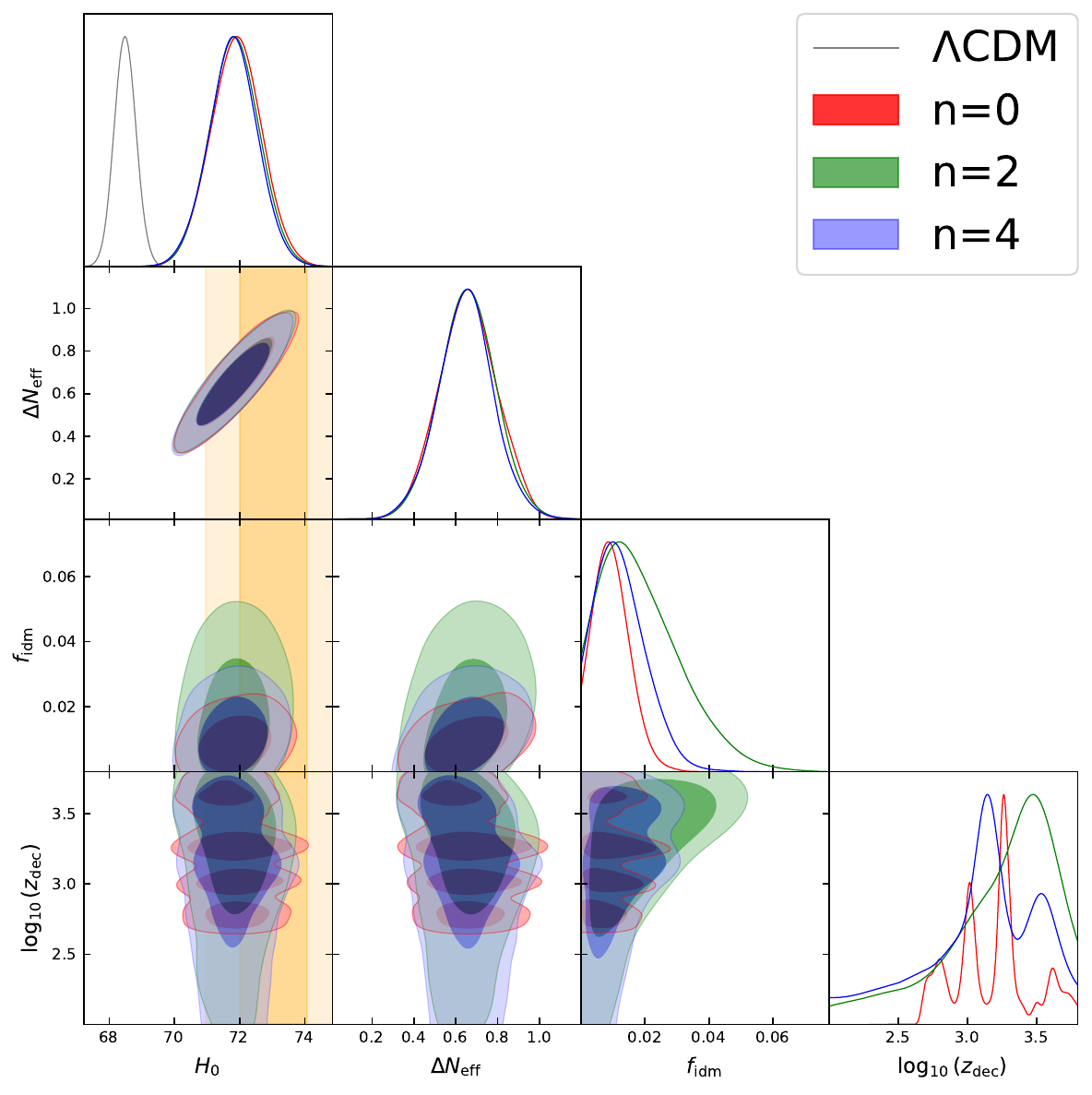}
	\caption{
    1D and 2D posterior distributions of the $H_0$, $\DNeff$, $f_\idm$, and $\log_{10} (z_{\dec})$ parameters of the \lcdm (gray), $n=0$ (red), $n=2$ (green), and $n=4$ (blue) models, fitted to the $\mPHF$ dataset. The orange bands show the $1\sigma$ and $2\sigma$ contours of the $H_0$ measurement by the SH0ES collaboration \cite{Riess:2021jrx}.
    }
	\label{fig:PHF-triangles}
\end{figure}

In Figs.~\ref{fig:P-triangles}, \ref{fig:PH-triangles}, \ref{fig:PF-triangles}, and \ref{fig:PHF-triangles}, we present the 2D posterior contours of the parameters $H_0$, $\DNeff$, $f_\idm$, and $\logzdec$ obtained from fitting the \lcdm, $n=0$, $n=2$, and $n=4$ models to the $\mP$, $\mPH$, $\mPF$, and $\mPHF$ datasets, respectively. One sees that in all the fits, except for $\mPF$, \DNeff and $f_\idm$ are correlated, confirming that the inclusion of DM-DR interactions allows for a larger amount of DR. We include the $1\sigma$ and $2\sigma$ bands of the $H_0$ measurement from the SH0ES collaboration used in our $\mH$ dataset. These plots illustrate that, even when the SH0ES data is included, the \lcdm model is tightly constrained by the CMB and cannot accommodate significantly higher values of $H_0$. They further show that, as previously discussed, the DR–iDM interactions can alleviate the Hubble tension.
In particular, the $2\sigma$ contour of the $n=0$ model fit to $\mP$ reaches the central value of the SH0ES measurement, while the fit of all \genidm models to $\mPH$ shows excellent agreement. Including the $\mF$ dataset reduces both the best-fit and the maximum allowed values of $H_0$, and slightly worsens the overall fit. Nevertheless, the model continues to provide an excellent description of the combined data, showing that models of interacting dark matter and dark radiation represent a promising framework to address the Hubble tension.
In addition, the fits that do not include full-shape information also show that large $\DNeff$ and $f_\idm$ correlate with $\logzdec \approx 3.3$, i.e. the decoupling of the dark interactions should happen close to matter-radiation equality. The results that include full shape information push the preference for the decoupling redshift to higher values in regions where $f_\idm$ is sizable, showing that BOSS data provides important input in these scenarios.

\begin{figure}[tbh!]
    \centering
	\includegraphics[width=0.75\linewidth]{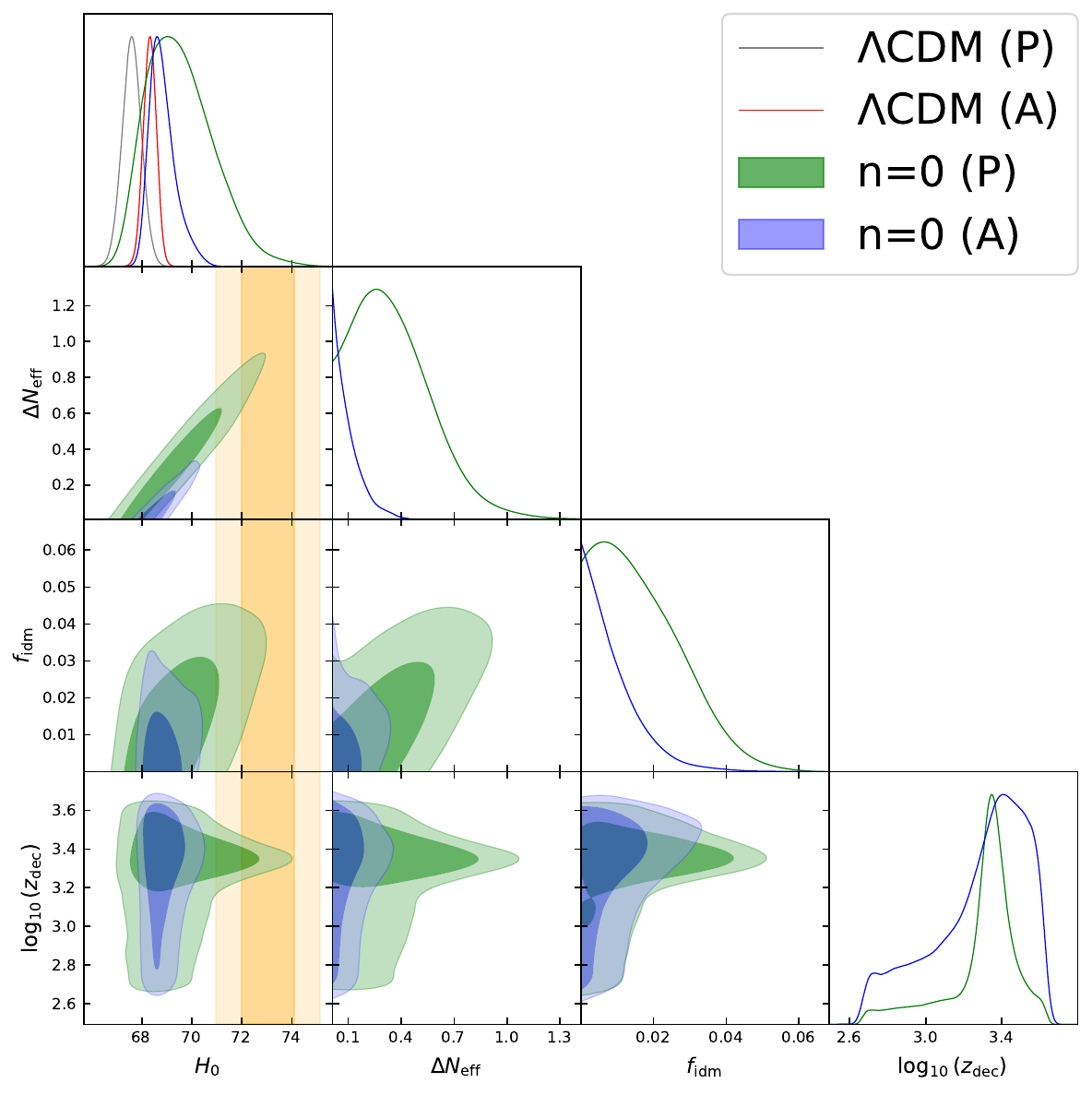}
	\caption{
    1D and 2D posterior distributions of the $H_0$, $\DNeff$, $f_\idm$, and $\log_{10} (z_{\dec})$ parameters of the \lcdm (gray and red), $n=0$ (green and blue) models, fitted to the $\mA$ dataset. The orange bands show the $1\sigma$ and $2\sigma$ contours of the $H_0$ measurement by the SH0ES collaboration \cite{Riess:2021jrx}.
    }
	\label{fig:A-triangles}
\end{figure}

\begin{figure}[tbh!]
    \centering
	\includegraphics[width=0.75\linewidth]{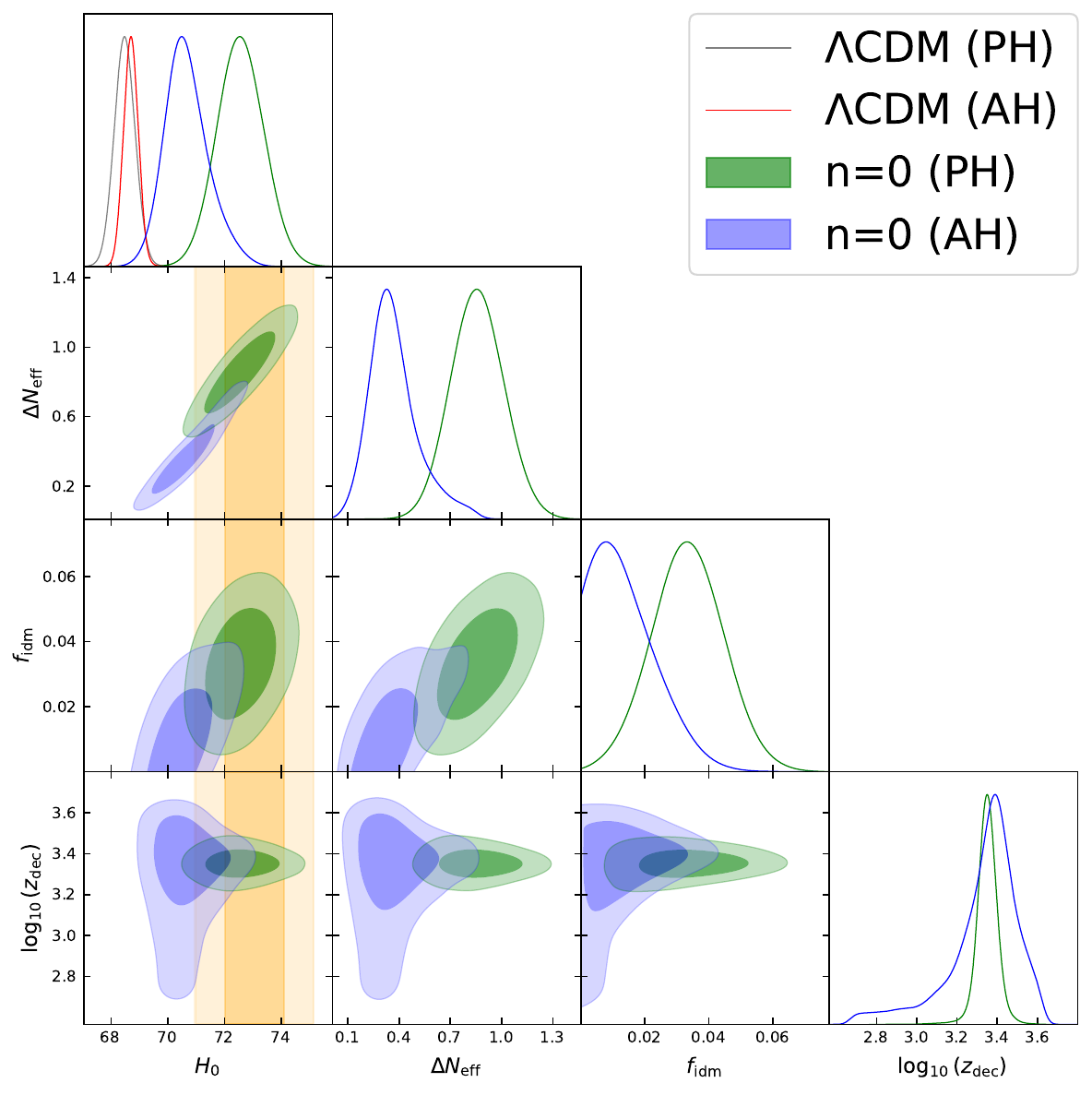}
	\caption{
    1D and 2D posterior distributions of the $H_0$, $\DNeff$, $f_\idm$, and $\log_{10} (z_{\dec})$ parameters of the \lcdm (gray and red), $n=0$ (green and blue) models, fitted to the $\mAH$ dataset. The orange bands show the $1\sigma$ and $2\sigma$ contours of the $H_0$ measurement by the SH0ES collaboration \cite{Riess:2021jrx}.
    }
	\label{fig:AH-triangles}
\end{figure}

Figs.~\ref{fig:A-triangles} and \ref{fig:AH-triangles} show the corresponding contours for the \lcdm and $n=0$ models fitted to the $\mP$ and $\mA$ datasets, and to the $\mPH$ and $\mAH$ datasets, respectively, allowing a direct comparison between the results obtained from the Planck and ACT data.
In the fit to $\mA$, the $n=0$ model yields a $2\sigma$ contour extending up to $H_0 \simeq 71~\km/\seg/\Mpc$, which, however, still falls short of the $2\sigma$ band of the SH0ES measurement. This is a reflection of the fact that the ACT data has tightened the constraints on the power spectra at high $\ell$, and shows no preference for an increase in diffusion damping. When the SH0ES dataset is included in the $\mAH$ fit, the inferred $H_0$ value is also lower than that obtained from $\mPH$, such that the $2\sigma$ contour of the $n=0$ model barely reaches the SH0ES central value. As discussed earlier, the ACT contribution to $\chi^2$ is larger for the interacting model best fit than for the \lcdm best fit, although the total $\Delta\chi^2$ for $\mAH$ remains negative because of the improved fit to SH0ES.

{\it In summary}, our analysis reveals several key findings regarding the \genidm models. First, the extended models are preferred over \lcdm only when the local SH0ES $H_0$ measurement is included in the fits, with the $n=0$ variant showing the strongest preference ($\Delta\text{AIC} \approx -29$ for $\mPH$). Without the SH0ES data, the penalty from additional degrees of freedom favors \lcdm. Second, the $n=0$ model demonstrates a remarkable ability to accommodate higher $H_0$ values while maintaining an excellent fit to the Planck data, effectively reducing the Hubble tension to less than $2 \, \sigma$. This is a clear indication that the abrupt decoupling in the $n=0$ case substantially modifies the sensitivity to DR constraints at the scales probed by Planck, allowing for larger values of $\DNeff$ even without the SH0ES data. Third, the inclusion of ACT data, which extends sensitivity to higher $\ell$, tightens constraints on \DNeff as this data disfavors the increased power suppression at small scales associated with higher DR, limiting the extent to which the model can resolve the Hubble tension. Finally, full-shape BOSS data provides additional constraints, favoring higher decoupling redshifts in regions of sizable $f_\idm$, while the correlation between $\DNeff$ and $f_\idm$ confirms that DM-DR interactions enable larger amounts of DR. Overall, we find that the $n=0$
model can substantially reduce the tension between the SH0ES result and Planck CMB data, but non-negligible residual tension remains when the analysis includes the matter power spectrum from BOSS or small-scale CMB measurements from ACT DR6.

\section{Conclusions}
\label{sec:conclusions}

In this work, we have performed a detailed phenomenological analysis of a class of cosmological models with interactions between DM and DR that have been proposed to address the Hubble tension. We explored three benchmark scenarios characterized by the temperature scaling of the momentum transfer rate, $\Gamma_d \propto T^{2+n}$, corresponding to four-fermion ($n=4$), Compton-like ($n=2$), and Coulomb-like ($n=0$) interactions, which differ in the rate of decoupling between the DR and DM.\footnote{
Our conclusions are based on the three physically motivated interaction-rate parametrizations considered here. A systematic exploration of alternative phenomenological forms is beyond the scope of this work.
} The iDM-DR interactions lead to a decrease of the MPS at small scales and introduce characteristic dark acoustic oscillations that are a striking signal of these models. The dark acoustic oscillations are more pronounced when the decoupling transition is faster. The iDM-DR interactions also impact the CMB through the change in time-evolution of the gravitational potential, and have the effect of relaxing the bounds on \DNeff at the scales probed by Planck.

Using a modified version of the {\tt CLASS} Boltzmann code to extend the ETHOS framework, we carried out MCMC analyses with recent cosmological datasets, including Planck 2018, ACT DR6, BAO, DESI, full-shape MPS, supernova, and local $H_0$ measurements. Without the SH0ES prior, all interacting models yield fits statistically comparable to \lcdm, indicating that current cosmological data alone do not yet prefer dark sector interactions. However, once the local $H_0$ measurement is included, all \genidm models provide significantly better fits, particularly the $n=0$ (Coulomb-like) case, which achieves an improvement of $\Delta\mathrm{AIC} \simeq -30$ for the $\mPH$ dataset. The preferred parameter space corresponds to a small iDM fraction, $f_\idm \simeq 1-3\%$, and a modest DR contribution, $\DNeff \simeq 0.7-0.9$. The resulting Hubble constant, $H_0 \simeq 72~\mathrm{\km/\seg/\Mpc}$, alleviates the Hubble tension while preserving the consistency of the Planck CMB spectra. The preference is substantially reduced when ACT DR6 data are included. The higher-precision small-scale anisotropy measurements from ACT constrain $\DNeff$ and the damping tail more tightly, reducing the statistical improvement seen in the $\mPH$ fits. Consequently, the best-fit values shift toward smaller $\DNeff$, thereby reducing the increase in $H_0$ to $H_0 \simeq 70~\mathrm{\km/\seg/\Mpc}$ for $\mAH$. 

Overall, our results highlight that a small interacting dark sector component can provide an elegant early-universe mechanism to reduce the Hubble tension, but the statistical preference depends sensitively on the inclusion of small-scale CMB data.  As seen in our results, full-shape analysis of large scale structure data and high precision measurements of small scale CMB data are already providing strong constraints on new physics proposals that can address the Hubble tension. The sensitivity of these probes is going to improve significantly over the next few years and will provide a crucial test of this class of interacting dark sector models.

\acknowledgments{
The authors thank Kimberly Boddy, Melissa Joseph, and Yuhsin Tsai for helpful discussions and feedback.
MBA, IF and ZC are supported in part by the National Science Foundation under Grant Number PHY-2514660, and the Maryland Center for Fundamental Physics.
The research of MBA was supported in part by grant no. NSF PHY-2309135 to the Kavli Institute for Theoretical Physics (KITP).
GMT is supported in part by the National Science Foundation under Grant Number PHY-2412828.
The research of CK is supported by the National Science Foundation Grant Number PHY-2210562.
TY is funded by the NSF grant PHY-2309456 and the Samsung Science and Technology Foundation under Project Number SSTF-BA2201-06.
The research of TY was supported by the Munich Institute for Astro-, Particle and BioPhysics (MIAPbP) which is funded by the Deutsche Forschungsgemeinschaft (DFG, German Research Foundation) under Germany´s Excellence Strategy – EXC-2094 – 390783311.
This work of TY was performed in part at Aspen Center for Physics, which is supported by NSF grant PHY-2210452 and a grant from the Simons Foundation (1161654, Troyer). 
}

\appendix

\section{Numerical Results}
\label{appendix}

In this appendix we summarize the results from our likelihood analysis.
The best fit values of the \lcdm and \genidm models to various combinations involving the datasets $\mP$, $\mH$, and $\mF$, discussed in \Sec{sec:meth}, are shown in Tables~\ref{tab:lcdm_val_bf} (\lcdm), \ref{tab:n0_val_bf} ($n=0$), \ref{tab:n2_val_bf} ($n=2$), and \ref{tab:n4_val_bf} ($n=4$).
The mean $\pm1\sigma$ values from the parameter posteriors are shown in
Tables~\ref{tab:lcdm_val_mn} (\lcdm), \ref{tab:n0_val_mn} ($n=0$), \ref{tab:n2_val_mn} ($n=2$), and \ref{tab:n4_val_mn} ($n=4$).
For the dataset combinations involving $\mA$ and $\mH$ instead, the best fit and mean $\pm 1 \sigma$ values are shown in Tables~\ref{tab:lcdm_val_bfmn_act} (\lcdm) and \ref{tab:n0_val_bfmn_act} ($n=0$).

\begin{table}[h!]
\centering
\begin{adjustbox}{max width=\columnwidth}
\begin{tabular}{@{}|c|cccc|@{}}
\toprule
\toprule
Parameter & $\mP$ & $\mPH$ & $\mPF$ & $\mPHF$ \\ \midrule
$100~\theta_s$  & $1.0418$ & $1.0422$ & $1.0420$ & $1.0421$ \\
$100~\omega_{b}$     & $2.243$ & $2.262$ & $2.251$ & $2.261$ \\
$\omega_\dm$       & $0.1192$ & $0.1173$ & $0.1184$ & $0.1180$ \\
$\ln 10^{10} A_s$        & $3.052$ & $3.068$ & $3.055$ & $3.049$ \\
$n_s$        & $0.9656$ & $0.9724$ & $0.9684$ & $0.9713$ \\
$\tau_\mathrm{reio}$        & $0.05705$ & $0.06504$ & $0.06184$ & $0.05807$ \\ \midrule
$M_B$        & $-19.429$ & $-19.401$ & $-19.419$ & $-19.410$ \\
$H_0~[\km/\seg/\Mpc]$        & $67.69$ & $68.68$ & $68.09$ & $68.36$ \\
$\sigma_8$    & $0.8112$ & $0.8127$ & $0.8106$ & $0.8074$ \\
$S_8$        & $0.8251$ & $0.8098$ & $0.8179$ & $0.8105$ \\ 
\bottomrule
\bottomrule
\end{tabular}
\end{adjustbox}
\caption{Best-fit values of the \lcdm model to various combinations of the $\mP$, $\mH$, and $\mF$ datasets.}
\label{tab:lcdm_val_bf}
\end{table}

\begin{table}[h!]
\centering
\begin{adjustbox}{max width=\columnwidth}
\begin{tabular}{@{}|c|cccc|@{}}
\toprule
\toprule
Parameter & $\mP$ & $\mPH$ & $\mPF$ & $\mPHF$ \\ \midrule
$100~\theta_s$  & $1.0431$ & $1.0435$ & $1.0421$ & $1.0426$ \\
$100~\omega_{b}$     & $2.273$ & $2.300$ & $2.248$ & $2.292$ \\
$\omega_\dm$       & $0.1278$ & $0.1379$ & $0.1209$ & $0.1306$ \\
$\ln 10^{10} A_s$        & $3.061$ & $3.036$ & $3.050$ & $3.040$ \\
$n_s$        & $0.9755$ & $0.9742$ & $0.9683$ & $0.9712$ \\
$\tau_\mathrm{reio}$        & $0.05888$ & $0.05424$ & $0.05688$ & $0.05546$ \\
$\Delta\Neff$       & $0.3760$ & $0.8914$ & $0.0551$ & $0.6170$ \\
$f_\idm [\%]$        & $2.6848$ & $2.8804$ & $1.0558$ & $0.7825$ \\
\logztr     & $4.686$ & $4.684$ & $4.945$ & $4.633$ \\
\midrule
\logzdec     & $3.333$ & $3.331$ & $3.589$ & $3.281$ \\
$M_B$        & $-19.365$ & $-19.282$ & $-19.424$ & $-19.312$ \\
$H_0~[\km/\seg/\Mpc]$        & $69.76$ & $72.61$ & $67.88$ & $71.57$ \\
$\sigma_8$    & $0.8079$ & $0.8110$ & $0.8065$ & $0.8187$ \\
$S_8$        & $0.8221$ & $0.8198$ & $0.8233$ & $0.8201$ \\ 
\bottomrule
\bottomrule
\end{tabular}
\end{adjustbox}
\caption{Best-fit values of the $n=0$ model to various combinations of the $\mP$, $\mH$, and $\mF$ datasets.}
\label{tab:n0_val_bf}
\end{table}

\begin{table}[h!]
\centering
\begin{adjustbox}{max width=\columnwidth}
\begin{tabular}{@{}|c|cccc|@{}}
\toprule
\toprule
Parameter & $\mP$ & $\mPH$ & $\mPF$ & $\mPHF$ \\ \midrule
$100~\theta_s$  & $1.0421$ & $1.0428$ & $1.0422$ & $1.0429$ \\
$100~\omega_{b}$     & $2.255$ & $2.301$ & $2.257$ & $2.300$ \\
$\omega_\dm$       & $0.1226$ & $0.1332$ & $0.1223$ & $0.1335$ \\
$\ln 10^{10} A_s$        & $3.066$ & $3.046$ & $3.052$ & $3.047$ \\
$n_s$        & $0.9712$ & $0.9727$ & $0.9718$ & $0.9747$ \\
$\tau_\mathrm{reio}$        & $0.06316$ & $0.05745$ & $0.05667$ & $0.05852$ \\
$\Delta\Neff$       & $0.1707$ & $0.7343$ & $0.1350$ & $0.7310$ \\
$f_\idm [\%]$        & $0.4321$ & $1.7552$ & $1.9535$ & $2.8864$ \\
\logzdec     & $3.251$ & $3.187$ & $3.522$ & $3.431$ \\
\midrule
$M_B$        & $-19.400$ & $-19.295$ & $-19.408$ & $-19.296$ \\
$H_0~[\km/\seg/\Mpc]$        & $68.69$ & $72.18$ & $68.44$ & $72.10$ \\
$\sigma_8$    & $0.8209$ & $0.8156$ & $0.8030$ & $0.8096$ \\
$S_8$        & $0.8331$ & $0.8171$ & $0.8173$ & $0.8126$ \\ 
\bottomrule
\bottomrule
\end{tabular}
\end{adjustbox}
\caption{Best-fit values of the $n=2$ model to various combinations of the $\mP$, $\mH$, and $\mF$ datasets.}
\label{tab:n2_val_bf}
\end{table}

\begin{table}[h!]
\centering
\begin{adjustbox}{max width=\columnwidth}
\begin{tabular}{@{}|c|cccc|@{}}
\toprule
\toprule
Parameter & $\mP$ & $\mPH$ & $\mPF$ & $\mPHF$ \\ \midrule
$100~\theta_s$  & $1.0422$ & $1.0432$ & $1.0421$ & $1.0429$ \\
$100~\omega_{b}$     & $2.255$ & $2.291$ & $2.247$ & $2.295$ \\
$\omega_\dm$       & $0.1237$ & $0.1359$ & $0.1206$ & $0.1324$ \\
$\ln 10^{10} A_s$        & $3.050$ & $3.049$ & $3.045$ & $3.033$ \\
$n_s$        & $0.9688$ & $0.9713$ & $0.9657$ & $0.9724$ \\
$\tau_\mathrm{reio}$        & $0.05570$ & $0.06050$ & $0.05416$ & $0.05634$ \\
$\Delta\Neff$       & $0.1819$ & $0.7909$ & $0.0573$ & $0.7144$ \\
$f_\idm [\%]$        & $0.9960$ & $2.7305$ & $0.0800$ & $1.2361$ \\
\logzdec     & $3.325$ & $3.254$ & $3.575$ & $3.1525$ \\
\midrule
$M_B$        & $-19.403$ & $-19.300$ & $-19.421$ & $-19.293$ \\
$H_0~[\km/\seg/\Mpc]$        & $68.47$ & $71.96$ & $67.99$ & $72.22$ \\
$\sigma_8$    & $0.8115$ & $0.8133$ & $0.8049$ & $0.8137$ \\
$S_8$        & $0.8294$ & $0.8240$ & $0.8193$ & $0.8124$ \\ 
\bottomrule
\bottomrule
\end{tabular}
\end{adjustbox}
\caption{Best-fit values of the $n=4$ model to various combinations of the $\mP$, $\mH$, and $\mF$ datasets.}
\label{tab:n4_val_bf}
\end{table}

\begin{table}[h!]
\centering
\begin{adjustbox}{max width=\columnwidth}
\begin{tabular}{@{}|c|cccc|@{}}
\toprule
\toprule
Parameter & $\mP$ & $\mPH$ & $\mPF$ & $\mPHF$ \\ \midrule
$100~\theta_s$  & $1.0419^{+0.0003}_{-0.0003}$ & $1.0421^{+0.0003}_{-0.0003}$ & $1.0419^{+0.0003}_{-0.0003}$ & $1.0421^{+0.0003}_{-0.0003}$ \\
$100~\omega_{b}$     & $2.240^{+0.013}_{-0.013}$ & $2.259^{+0.013}_{-0.013}$ & $2.243^{+0.013}_{-0.013}$ & $2.260^{+0.012}_{-0.012}$ \\
$\omega_\dm$       & $0.1195^{+0.0009}_{-0.0009}$ & $0.1177^{+0.0008}_{-0.0008}$ & $0.1191^{+0.0008}_{-0.0008}$ & $0.1177^{+0.0007}_{-0.0007}$ \\
$\ln 10^{10} A_s$        & $3.047^{+0.013}_{-0.013}$ & $3.056^{+0.012}_{-0.014}$ & $3.050^{+0.013}_{-0.014}$ & $3.056^{+0.014}_{-0.014}$ \\
$n_s$        & $0.9660^{+0.0036}_{-0.0036}$ & $0.9705^{+0.0035}_{-0.0035}$ & $0.9666^{+0.0035}_{-0.0035}$ & $0.9703^{+0.0034}_{-0.0034}$ \\
$\tau_\mathrm{reio}$        & $0.05625^{+0.00669}_{-0.00675}$ & $0.06191^{+0.00630}_{-0.00724}$ & $0.05789^{+0.00647}_{-0.00723}$ & $0.06202^{+0.00663}_{-0.00742}$ \\ \midrule
$M_B$        & $-19.434^{+0.011}_{-0.011}$ & $-19.406^{+0.010}_{-0.010}$ & $-19.428^{+0.010}_{-0.010}$ & $-19.406^{+0.010}_{-0.010}$ \\
$H_0~[\km/\seg/\Mpc]$        & $67.60^{+0.38}_{-0.38}$ & $68.48^{+0.36}_{-0.36}$ & $67.78^{+0.34}_{-0.35}$ & $68.49^{+0.33}_{-0.33}$ \\
$\sigma_8$    & $0.8107^{+0.0055}_{-0.0055}$ & $0.8088^{+0.0053}_{-0.0059}$ & $0.8106^{+0.0057}_{-0.0057}$ & $0.8088^{+0.0058}_{-0.0058}$ \\
$S_8$        & $0.8266^{+0.0099}_{-0.0099}$ & $0.8095^{+0.0095}_{-0.0094}$ & $0.8232^{+0.0091}_{-0.0091}$ & $0.8094^{+0.0088}_{-0.0088}$ \\ 
\bottomrule
\bottomrule
\end{tabular}
\end{adjustbox}
\caption{Mean $\pm 1\sigma$ values of the \lcdm model to various combinations of the $\mP$, $\mH$, and $\mF$ datasets.}
\label{tab:lcdm_val_mn}
\end{table}

\begin{table}[h!]
\centering
\begin{adjustbox}{max width=\columnwidth}
\begin{tabular}{@{}|c|cccc|@{}}
\toprule
\toprule
Parameter & $\mP$ & $\mPH$ & $\mPF$ & $\mPHF$ \\ \midrule
$100~\theta_s$  & $1.0426^{+0.0004}_{-0.0006}$ & $1.0435^{+0.0005}_{-0.0005}$ & $1.0423^{+0.0003}_{-0.0004}$ & $1.0428^{+0.0004}_{-0.0004}$ \\
$100~\omega_{b}$ & $2.268^{+0.020}_{-0.025}$ & $2.312^{+0.017}_{-0.017}$ & $2.263^{+0.016}_{-0.019}$ & $2.292^{+0.015}_{-0.015}$ \\
$\omega_\dm$ & $0.1270^{+0.0031}_{-0.0058}$ & $0.1374^{+0.0038}_{-0.0038}$ & $0.1239^{+0.0024}_{-0.0040}$ & $0.1314^{+0.0028}_{-0.0034}$ \\
$\ln 10^{10} A_s$ & $3.049^{+0.014}_{-0.014}$ & $3.050^{+0.015}_{-0.015}$ & $3.049^{+0.011}_{-0.012}$ & $3.046^{+0.012}_{-0.011}$ \\
$n_s$ & $0.9710^{+0.0044}_{-0.0053}$ & $0.9776^{+0.0045}_{-0.0045}$ & $ 0.9690^{+0.0038}_{-0.0038}$ & $0.9723^{+0.0039}_{-0.0039}$ \\
$\tau_\mathrm{reio}$ & $0.05718^{+0.00656}_{-0.00717}$ & $0.05892^{+0.00676}_{-0.00764}$ & $0.05849^{+0.00587}_{-0.00615}$ & $0.06045^{+0.00615}_{-0.00608}$ \\
$\Delta\Neff$ & $0.3497^{+0.1287}_{-0.3029}$ & $0.8592^{+0.1554}_{-0.1549}$ & $0.2451^{+0.0714}_{-0.2351}$ & $0.6638^{+0.1437}_{-0.1399}$ \\
$f_\idm [\%]$ & $1.6883^{+0.4657}_{-1.6883}$ & $3.3365^{+1.1235}_{-1.1252}$ & $0.8967^{+0.1884}_{-0.8967}$ & $1.0129^{+0.4434}_{-0.6838}$ \\
\logztr & $4.649^{+0.198}_{-0.062}$ & $4.703^{+0.047}_{-0.045}$ & $4.723^{+0.467}_{-0.392}$ & $4.536^{+0.460}_{-0.462}$ \\
\midrule
\logzdec & $3.296^{+0.194}_{-0.060}$ & $3.350^{+0.047}_{-0.044}$ & $3.369^{+0.435}_{-0.396}$ & $3.185^{+0.452}_{-0.457}$ \\
$M_B$ & $-19.373^{+0.030}_{-0.050}$ & $-19.281^{+0.024}_{-0.024}$ & $-19.383^{+0.024}_{-0.041}$ & $-19.302^{+0.022}_{-0.023}$ \\
$H_0~[\km/\seg/\Mpc]$ & $69.55^{+0.98}_{-1.63}$ & $72.56^{+0.82}_{-0.82}$ & $69.24^{+0.78}_{-1.36}$ & $71.92^{+0.77}_{-0.79}$ \\
$\sigma_8$ & $0.8096^{+0.0080}_{-0.0072}$ & $0.8118^{+0.0076}_{-0.0076}$ & $0.8107^{+0.0087}_{-0.0068}$ & $0.8191^{+0.0067}_{-0.0060}$ \\
$S_8$ & $0.8239^{+0.0103}_{-0.0102}$ & $0.8199^{+0.0098}_{-0.0097}$ & $0.8200^{+0.0098}_{-0.0096}$ & $0.8184^{+0.0091}_{-0.0091}$ \\
\bottomrule
\bottomrule
\end{tabular}
\end{adjustbox}
\caption{Mean $\pm 1\sigma$ values of the $n=0$ model to various combinations of the $\mP$, $\mH$, and $\mF$ datasets.}
\label{tab:n0_val_mn}
\end{table}

\begin{table}[h!]
\centering
\begin{adjustbox}{max width=\columnwidth}
\begin{tabular}{@{}|c|cccc|@{}}
\toprule
\toprule
Parameter & $\mP$ & $\mPH$ & $\mPF$ & $\mPHF$ \\ \midrule
$100~\theta_s$  & $1.0423^{+0.0003}_{-0.0004}$ & $1.0429^{+0.0004}_{-0.0004}$ & $1.0423^{+0.0003}_{-0.0004}$ & $1.0428^{+0.0004}_{-0.0004}$ \\
$100~\omega_{b}$ & $2.256^{+0.016}_{-0.019}$ & $2.294^{+0.016}_{-0.016}$ &  $2.260^{+0.016}_{-0.018}$ & $2.295^{+0.015}_{-0.015}$ \\
$\omega_\dm$ & $0.1243^{+0.0020}_{-0.0039}$ & $0.1327^{+0.0031}_{-0.0035}$ & $0.1236^{+0.0019}_{-0.0035}$ & $0.1315^{+0.0029}_{-0.0030}$ \\
$\ln 10^{10} A_s$ & $3.048^{+0.014}_{-0.014}$ & $3.047^{+0.015}_{-0.016}$ & $3.050^{+0.014}_{-0.015}$ & $3.045^{+0.012}_{-0.012}$ \\
$n_s$ & $0.9688^{+0.0038}_{-0.0042}$ & $0.9739^{+0.0045}_{-0.0045}$ & $0.9693^{+0.0038}_{-0.0039}$ & $0.9733^{+0.0042}_{-0.0041}$ \\
$\tau_\mathrm{reio}$ & $0.05753^{+0.00664}_{-0.00731}$ & $0.06001^{+0.00687}_{-0.00801}$ & $0.05853^{+0.00696}_{-0.00698}$ & $0.05969^{+0.00609}_{-0.00615}$ \\
$\Delta\Neff$ & $0.2380^{+0.0733}_{-0.2162}$ & $0.7045^{+0.1416}_{-0.1426}$ & $0.2216^{+0.0563}_{-0.2116}$ & $0.6601^{+0.1292}_{-0.1312}$ \\
$f_\idm [\%]$ & $1.1486^{+0.1865}_{-1.1486}$ & $1.8415^{+0.5497}_{-1.7043}$ & $1.4489^{+0.3142}_{-1.4489}$ & $1.9054^{+0.6487}_{-1.6389}$ \\
\logzdec & $3.128^{+0.672}_{-0.163}$ & $3.115^{+0.426}_{-0.207}$ & $3.273^{+0.527}_{-0.087}$ & $3.209^{+0.515}_{-0.174}$ \\
\midrule
$M_B$ & $-19.391^{+0.023}_{-0.038}$ & $-19.299^{+0.023}_{-0.023}$ & $-19.389^{+0.021}_{-0.035}$ & $-19.304^{+0.022}_{-0.022}$ \\
$H_0~[\km/\seg/\Mpc]$ & $68.97^{+0.75}_{-1.24}$ & $71.99^{+0.79}_{-0.78}$ & $69.05^{+0.69}_{-1.15}$ & $71.86^{+0.76}_{-0.76}$ \\
$\sigma_8$ & $0.8092^{+0.0090}_{-0.0071}$ & $0.8147^{+0.0094}_{-0.0077}$ & $0.8067^{+0.0100}_{-0.0073}$ & $0.8125^{+0.0097}_{-0.0074}$ \\
$S_8$ & $0.8226^{+0.0118}_{-0.0102}$ & $0.8167^{+0.0106}_{-0.0106}$ & $0.8173^{+0.0116}_{-0.0099}$ & $0.8130^{+0.0107}_{-0.0096}$ \\
\bottomrule
\bottomrule
\end{tabular}
\end{adjustbox}
\caption{Mean $\pm 1\sigma$ values of the $n=2$ model to various combinations of the $\mP$, $\mH$, and $\mF$ datasets.}
\label{tab:n2_val_mn}
\end{table}

\begin{table}[h!]
\centering
\begin{adjustbox}{max width=\columnwidth}
\begin{tabular}{@{}|c|cccc|@{}}
\toprule
\toprule
Parameter & $\mP$ & $\mPH$ & $\mPF$ & $\mPHF$ \\ \midrule
$100~\theta_s$  & $1.0424^{+0.0003}_{-0.0004}$ & $1.0431^{+0.0005}_{-0.0005}$ & $1.0423^{+0.0003}_{-0.0004}$ & $1.0428^{+0.0004}_{-0.0004}$ \\
$100~\omega_{b}$ & $2.259^{+0.017}_{-0.020}$ & $2.300^{+0.017}_{-0.017}$ & $2.261^{+0.016}_{-0.018}$ & $2.292^{+0.015}_{-0.015}$ \\
$\omega_\dm$ & $0.1247^{+0.0021}_{-0.0042}$ & $0.1342^{+0.0036}_{-0.0036}$ & $0.1234^{+0.0018}_{-0.0033}$ & $ 0.1312^{+0.0026}_{-0.0027}$ \\
$\ln 10^{10} A_s$ & $3.049^{+0.014}_{-0.014}$ & $3.048^{+0.014}_{-0.016}$ & $3.050^{+0.013}_{-0.014}$ & $3.046^{+0.010}_{-0.013}$ \\
$n_s$ & $0.9693^{+0.0040}_{-0.0044}$ & $0.9754^{+0.0048}_{-0.0048}$ & $0.9693^{+0.0038}_{-0.0038}$ & $0.9729^{+0.0039}_{-0.0045}$ \\
$\tau_\mathrm{reio}$ & $0.05756^{+0.00684}_{-0.00696}$ & $0.05989^{+0.00673}_{-0.00774}$ & $0.05869^{+0.00652}_{-0.00711}$ & $0.06007^{+0.00528}_{-0.00679}$ \\
$\Delta\Neff$ & $0.2534^{+0.0606}_{-0.2434}$ & $0.7519^{+0.1463}_{-0.1481}$ & $0.2152^{+0.0527}_{-0.2052}$ & $0.6482^{+0.1159}_{-0.1220}$ \\
$f_\idm [\%]$ & $1.1481^{+0.2255}_{-1.1481}$ & $2.3277^{+1.0444}_{-1.6545}$ & $1.1694^{+0.2322}_{-1.1694}$ & $1.2928^{+0.5257}_{-0.9959}$ \\
\logzdec & $3.128^{+0.650}_{-0.116}$ & $3.189^{+0.236}_{-0.055}$ & $3.237^{+0.563}_{-0.131}$ & $3.097^{+0.556}_{-0.178}$ \\
\midrule
$M_B$ & $-19.388^{+0.023}_{-0.039}$ & $-19.294^{+0.023}_{-0.023}$ & $-19.389^{+0.021}_{-0.035}$ & $-19.305^{+0.021}_{-0.021}$ \\
$H_0~[\km/\seg/\Mpc]$ & $69.05^{+0.75}_{-1.29}$ & $72.16^{+0.79}_{-0.79}$ & $69.05^{+0.68}_{-1.16}$ & $71.80^{+0.71}_{-0.72}$ \\
$\sigma_8$ & $0.8099^{+0.0085}_{-0.0069}$ & $0.8142^{+0.0089}_{-0.0080}$ & $0.8082^{+0.0093}_{-0.0067}$ & $0.8165^{+0.0069}_{-0.0068}$ \\
$S_8$ & $0.8235^{+0.0112}_{-0.0101}$ & $0.8184^{+0.0102}_{-0.0102}$ & $0.8184^{+0.0114}_{-0.0096}$ & $0.8168^{+0.0091}_{-0.0091}$ \\
\bottomrule
\bottomrule
\end{tabular}
\end{adjustbox}
\caption{Mean $\pm 1\sigma$ values of the $n=4$ model to various combinations of the $\mP$, $\mH$, and $\mF$ datasets.}
\label{tab:n4_val_mn}
\end{table}

\begin{table}[h!]
\centering
\begin{adjustbox}{max width=\columnwidth}
\begin{tabular}{@{}|c|cccc|@{}}
\toprule
\toprule
\multicolumn{1}{|c|}{\multirow{2}{*}{Parameter}} & \multicolumn{2}{c}{$\mA$}                       & \multicolumn{2}{c|}{$\mAH$}                        \\ 
\multicolumn{1}{|c|}{}                           & \multicolumn{1}{c}{Best Fit} & \multicolumn{1}{c}{Mean$\pm1\sigma$} & \multicolumn{1}{c}{Best Fit} & \multicolumn{1}{c|}{Mean$\pm1\sigma$} \\
\midrule
$100~\theta_s$  & $1.0418$ &  $1.0418^{+0.0002}_{-0.0002}$  & $1.0418$ & $1.0419^{+0.0002}_{-0.0003}$ \\
$100~\omega_{b}$     & $2.258$ & $2.256^{+0.010}_{-0.010}$ & $2.265$ & $2.264^{+0.010}_{-0.011}$ \\
$\omega_\dm$       & $0.1176$ & $0.1177^{+0.0006}_{-0.0006}$ & $0.1169$ & $0.1169^{+0.0006}_{-0.0006}$ \\
$\ln 10^{10} A_s$        & $3.062$ & $3.061^{+0.011}_{-0.012}$ & $3.067$ & $3.066^{+0.011}_{-0.012}$ \\
$n_s$        & $0.9755$ & $0.9746^{+0.0034}_{-0.0034}$ & $0.9760$ & $ 0.9765^{+0.0034}_{-0.0034}$ \\
$\tau_\mathrm{reio}$        & $0.06529$ & $0.06476^{+0.00553}_{-0.00631}$ & $0.06807$ & $0.06778^{+0.00577}_{-0.00649}$ \\
\midrule
$M_B$        & $-19.413$ & $-19.414^{+0.008}_{-0.008}$ & $-19.401$ & $-19.400^{+0.008}_{-0.008}$ \\
$H_0~[\km/\seg/\Mpc]$        & $68.35$ & $68.32^{+0.26}_{-0.26}$ & $68.69$ & $68.70^{+0.26}_{-0.26}$ \\
$\sigma_8$    & $0.8124$ & $0.8121^{+0.0045}_{-0.0046}$ & $0.8118$ & $0.8116^{+0.0046}_{-0.0046}$  \\
$S_8$        & $0.8144$ & $0.8146^{+0.0068}_{-0.0068}$ & $0.8080$ & $0.8075^{+0.0067}_{-0.0067}$ \\ 
\bottomrule
\bottomrule
\end{tabular}
\end{adjustbox}
\caption{Best-fit and mean $\pm 1\sigma$ values of the \lcdm model parameters to various combinations of the $\mA$ and $\mH$ datasets.}
\label{tab:lcdm_val_bfmn_act}
\end{table}

\begin{table}[h!]
\centering
\begin{adjustbox}{max width=\columnwidth}
\begin{tabular}{@{}|c|cccc|@{}}
\toprule
\toprule
\multicolumn{1}{|c|}{\multirow{2}{*}{Parameter}} & \multicolumn{2}{c}{$\mA$}                       & \multicolumn{2}{c|}{$\mAH$}                        \\ 
\multicolumn{1}{|c|}{}                           & \multicolumn{1}{c}{Best Fit} & \multicolumn{1}{c}{Mean$\pm1\sigma$} & \multicolumn{1}{c}{Best Fit} & \multicolumn{1}{c|}{Mean$\pm1\sigma$} \\
\midrule
$100~\theta_s$  & $1.0418$ &  $1.0420^{+0.0003}_{-0.0003}$  & $1.0423$ & $1.0424^{+0.0003}_{-0.0004}$ \\
$100~\omega_{b}$     & $2.260$ & $2.266^{+0.011}_{-0.014}$ & $2.294$ & $2.297^{+0.015}_{-0.019}$ \\
$\omega_\dm$       & $0.1179$ & $0.1199^{+0.0008}_{-0.0020}$ & $0.1239$ & $0.1251^{+0.0020}_{-0.0040}$ \\
$\ln 10^{10} A_s$        & $3.065$ & $3.062^{+0.011}_{-0.013}$ & $3.064$ & $3.061^{+0.009}_{-0.014}$ \\
$n_s$        & $0.9752$ & $0.9773^{+0.0036}_{-0.0040}$ & $0.9818$ & $0.9823^{+0.0043}_{-0.0048}$ \\
$\tau_\mathrm{reio}$        & $0.06666$ & $0.06467^{+0.00564}_{-0.00670}$ & $0.06746$ & $0.06588^{+0.00530}_{-0.00702}$ \\
$\Delta\Neff$       & $0.0185$ & $0.0976^{+0.0171}_{-0.0876}$ & $0.3285$ & $0.3780^{+0.0966}_{-0.1717}$ \\
$f_\idm [\%]$        & $0.0066$ & $0.8929^{+0.1646}_{-0.8929}$ & $1.3143$ & $1.4692^{+0.3618}_{-1.4692}$ \\
\logztr     & $4.743$ & $4.639^{+0.361}_{-0.071}$ & $4.741$ & $4.686^{+0.188}_{-0.090}$ \\
\midrule
\logzdec     & $3.389$ & $3.287^{+0.330}_{-0.095}$ & $3.388$ & $3.332^{+0.184}_{-0.088}$ \\
$M_B$        & $-19.409$ & $-19.397^{+0.011}_{-0.019}$ & $-19.346$ & $-19.339^{+0.020}_{-0.027}$ \\
$H_0~[\km/\seg/\Mpc]$        & $68.48$ & $68.84^{+0.34}_{-0.63}$ & $70.47$ & $70.70^{+0.65}_{-0.88}$ \\
$\sigma_8$    & $0.8133$ & $0.8087^{+0.0074}_{-0.0053}$ & $0.8132$ & $0.8125^{+0.0068}_{-0.0060}$  \\
$S_8$        & $0.8146$ & $0.8116^{+0.0081}_{-0.0070}$ & $0.8091$ & $0.8091^{+0.0072}_{-0.0072}$ \\ 
\bottomrule
\bottomrule
\end{tabular}
\end{adjustbox}
\caption{Best-fit and mean $\pm 1\sigma$ values of the $n=0$ model parameters to various combinations of the $\mA$ and $\mH$ datasets.}
\label{tab:n0_val_bfmn_act}
\end{table}

\begin{figure}[tbh!]
    \advance\leftskip-4pc
	\includegraphics[width=1.25\linewidth]{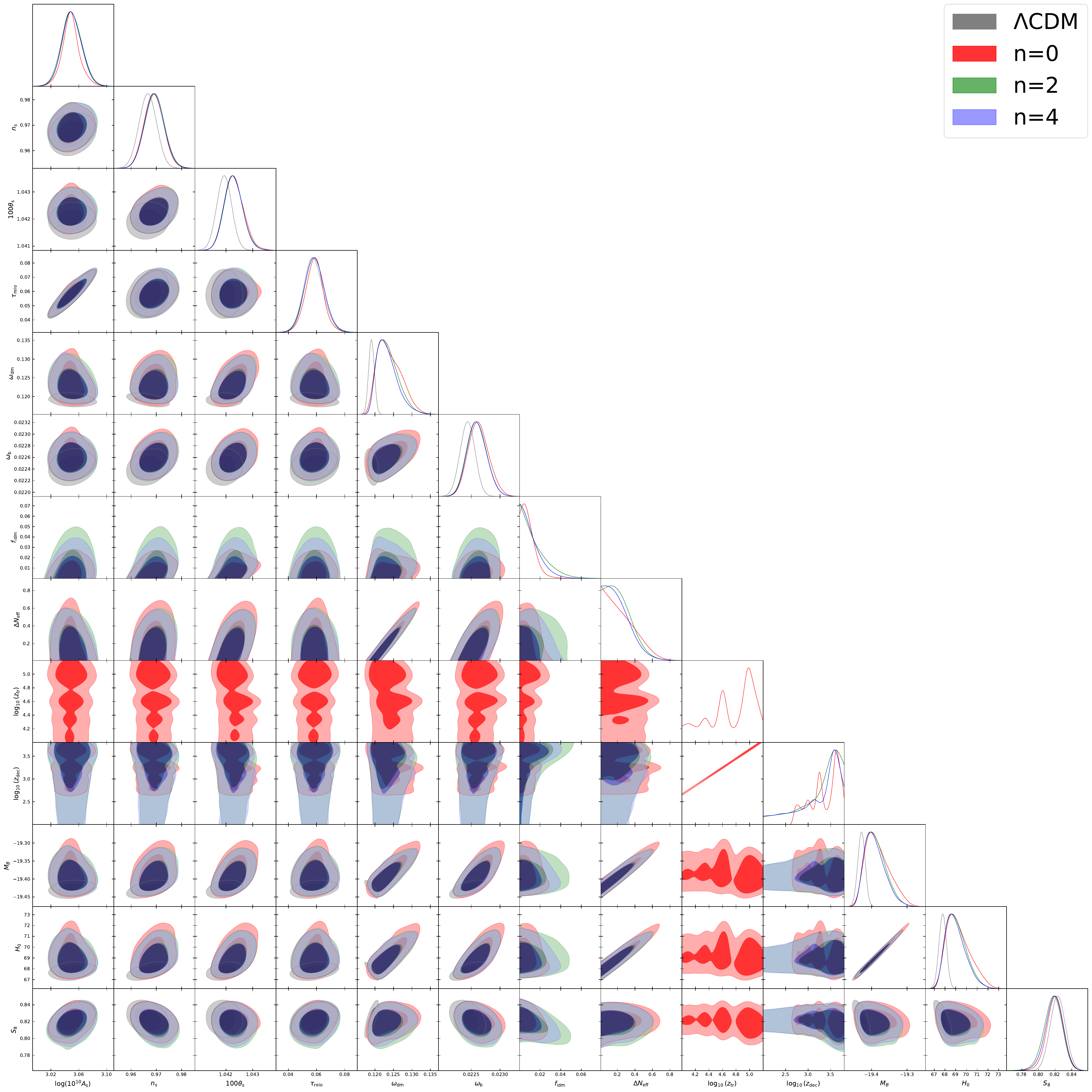}
	\caption{
    1D and 2D posterior distributions of the parameters of the \lcdm (gray), $n=0$ (red), $n=2$ (green), and $n=4$ (blue) models, fitted to the $\mPF$ datasets.
    }
	\label{fig:PF-triangle}
\end{figure}

\begin{figure}[tbh!]
    \advance\leftskip-4pc
	\includegraphics[width=1.25\linewidth]{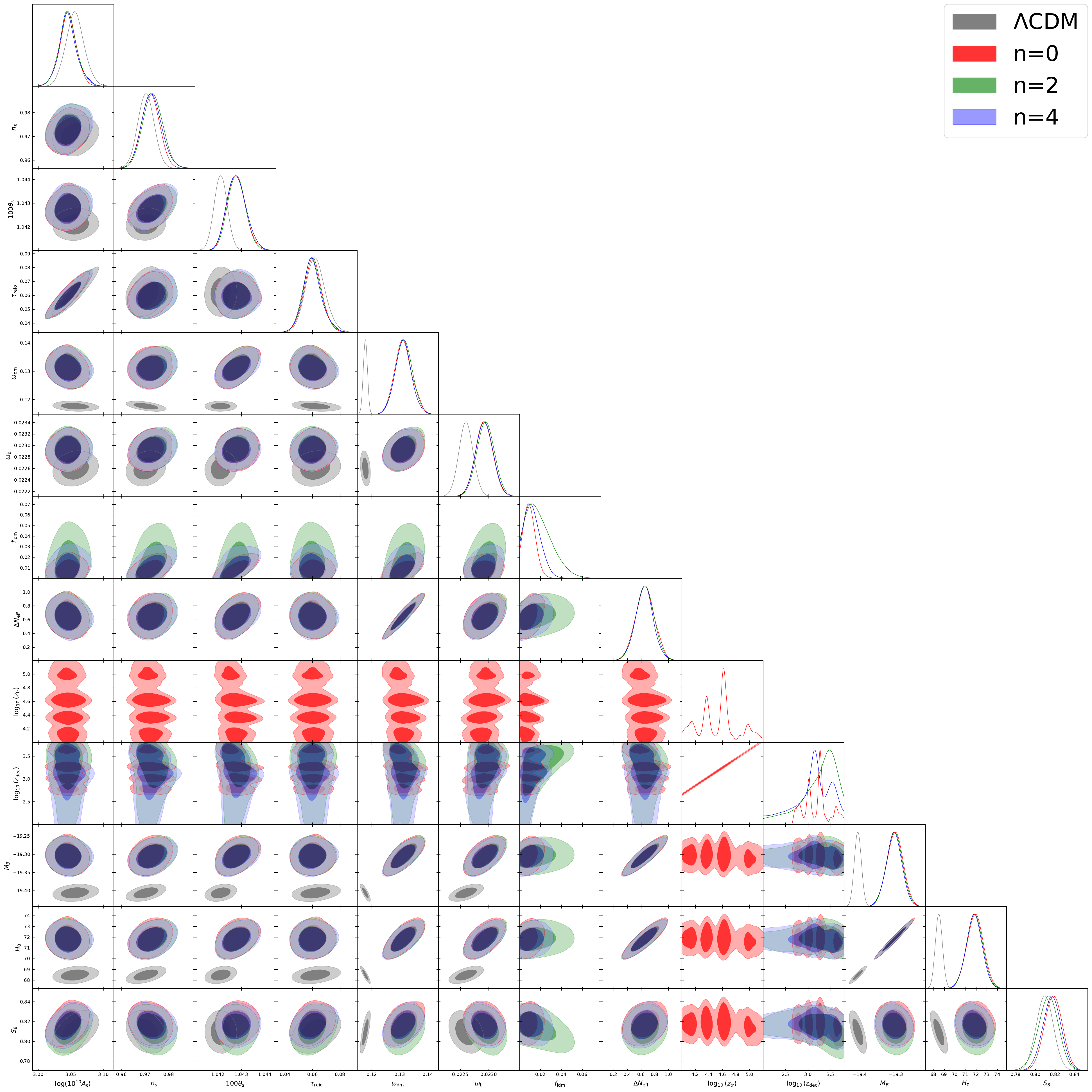}
	\caption{
    1D and 2D posterior distributions of the parameters of the \lcdm (gray), $n=0$ (red), $n=2$ (green), and $n=4$ (blue) models, fitted to the $\mPHF$ datasets.
    }
	\label{fig:PHF-triangle}
\end{figure}

\begin{figure}[tbh!]
    \advance\leftskip-4pc
	\includegraphics[width=1.25\linewidth]{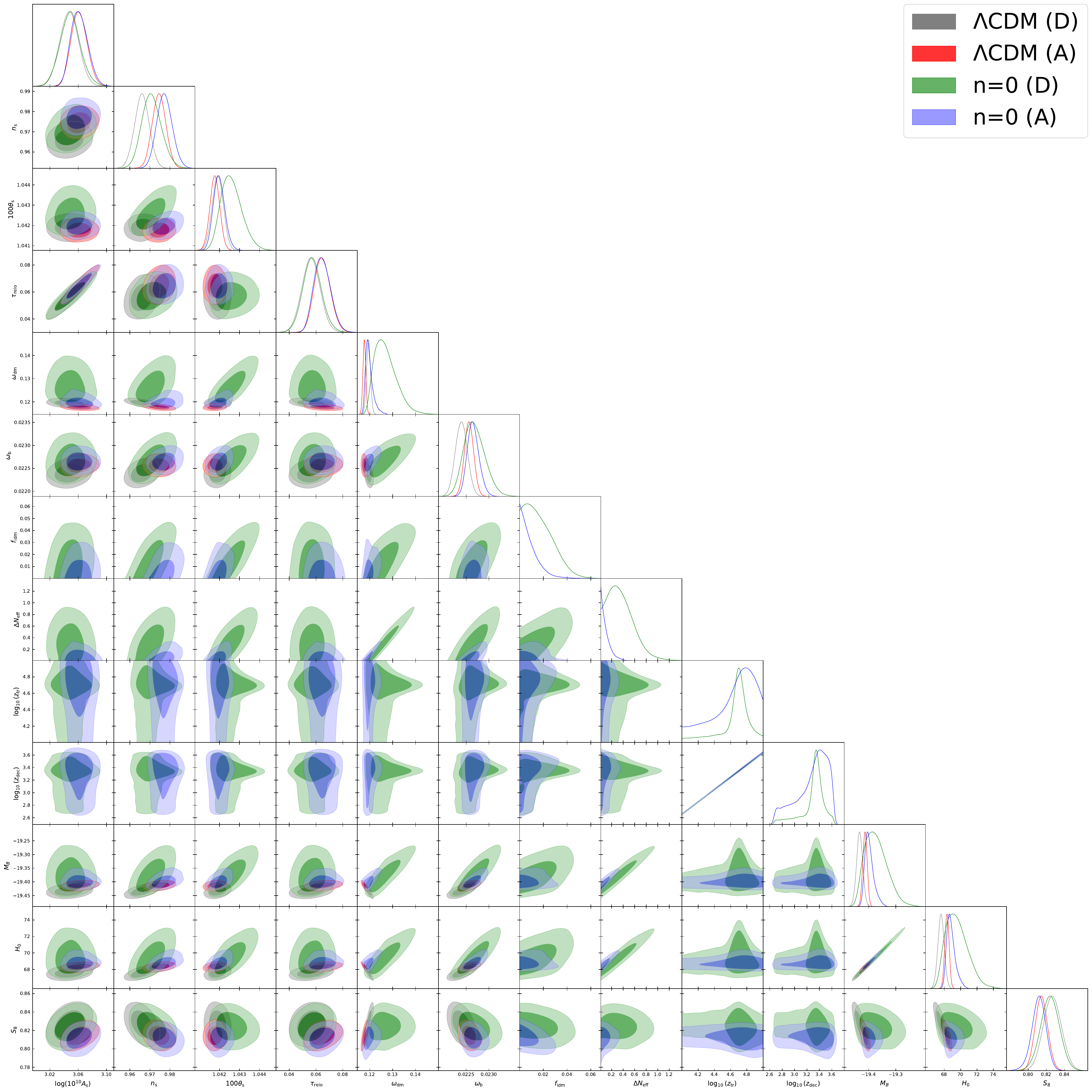}
	\caption{
    1D and 2D posterior distributions of the parameters of the \lcdm (gray and red) and $n=0$ (green and blue) models, fitted to the $\mP$ and $\mA$ datasets.
    }
	\label{fig:A-triangle}
\end{figure}

\begin{figure}[tbh!]
    \advance\leftskip-4pc
	\includegraphics[width=1.25\linewidth]{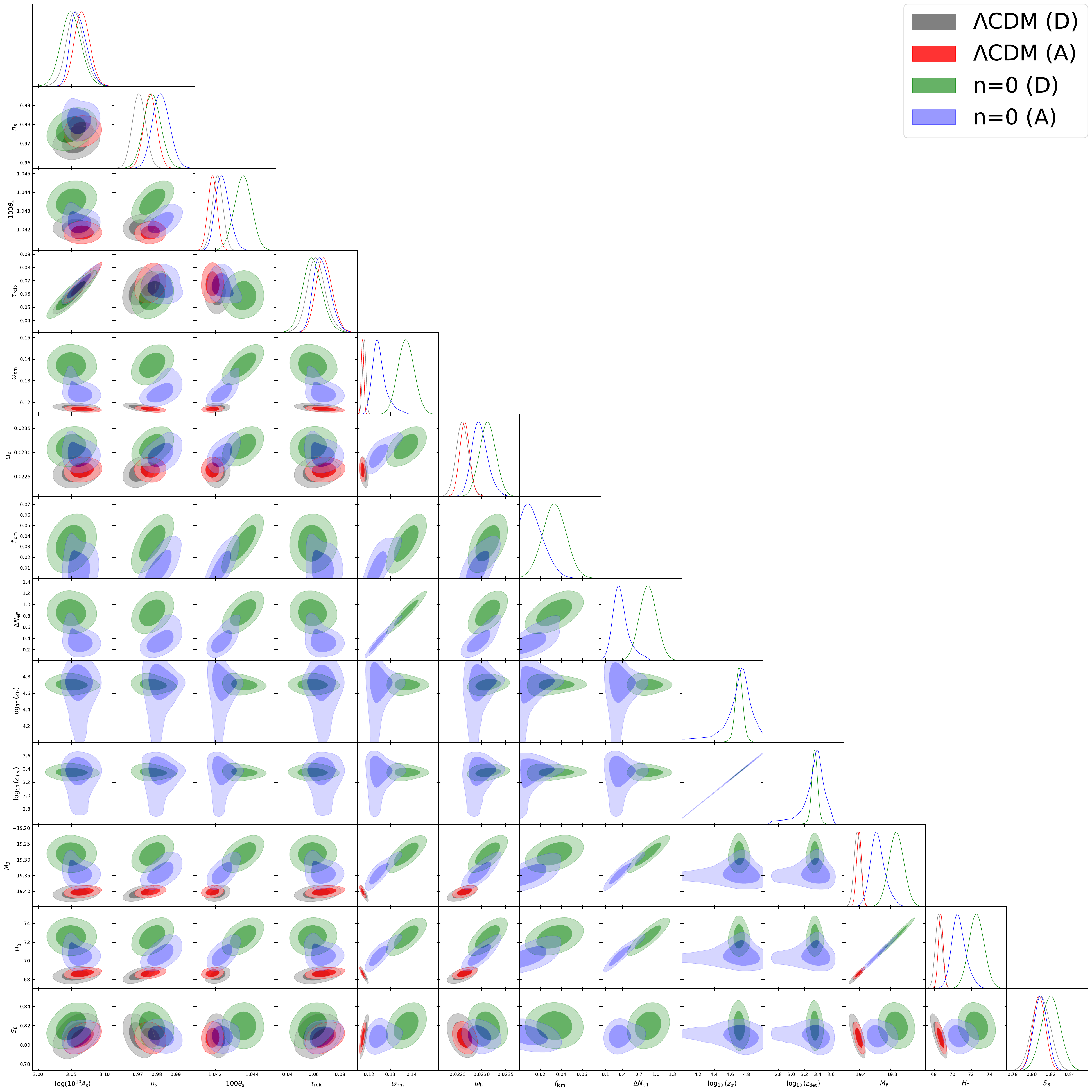}
	\caption{
    1D and 2D posterior distributions of the parameters of the \lcdm (gray and red) and $n=0$ (green and blue) models, fitted to the $\mPH$ and $\mAH$ datasets.
    }
	\label{fig:AH-triangle}
\end{figure}

In the following figures (\ref{fig:D-triangle}, \ref{fig:PH-triangle}, \ref{fig:PF-triangle}, \ref{fig:PHF-triangle}, \ref{fig:A-triangle}, and \ref{fig:AH-triangle}, for the $\mP$, $\mPH$, $\mPF$, $\mPHF$, $\mA$, and $\mAH$ datasets, respectively) we show the triangle plots (1D and 2D posterior distributions; $1\sigma$ and $2\sigma$ contours) of the most important parameters of the \lcdm, and $n=0$, $n=2$, and $n=4$ \genidm models.

\begin{figure}[tbh!]
    \advance\leftskip-4pc
	\includegraphics[width=1.25\linewidth]{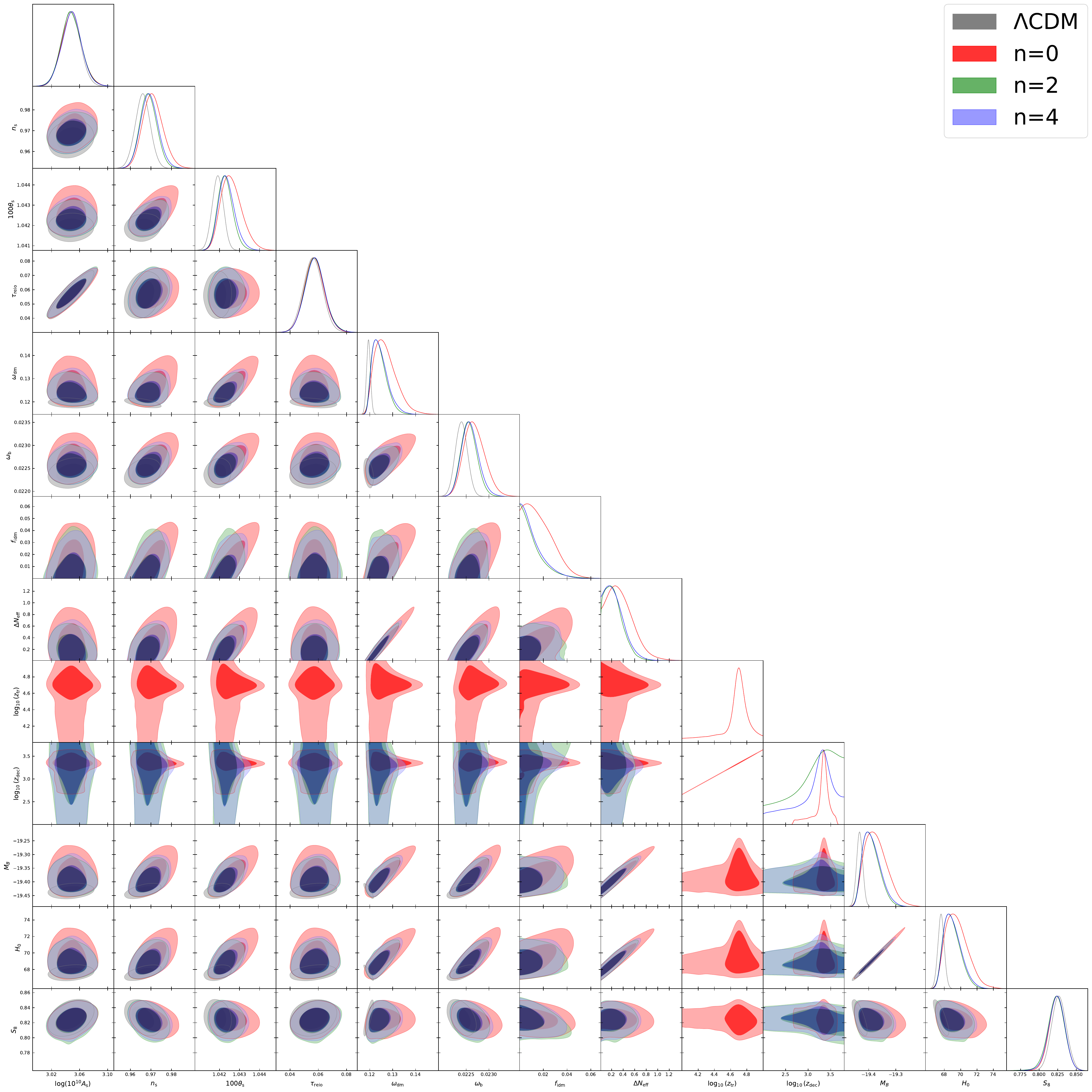}
	\caption{
    1D and 2D posterior distributions of the parameters of the \lcdm (gray), $n=0$ (red), $n=2$ (green), and $n=4$ (blue) models, fitted to the $\mP$ datasets.
    }
	\label{fig:D-triangle}
\end{figure}

\begin{figure}[tbh!]
    \advance\leftskip-4pc
	\includegraphics[width=1.25\linewidth]{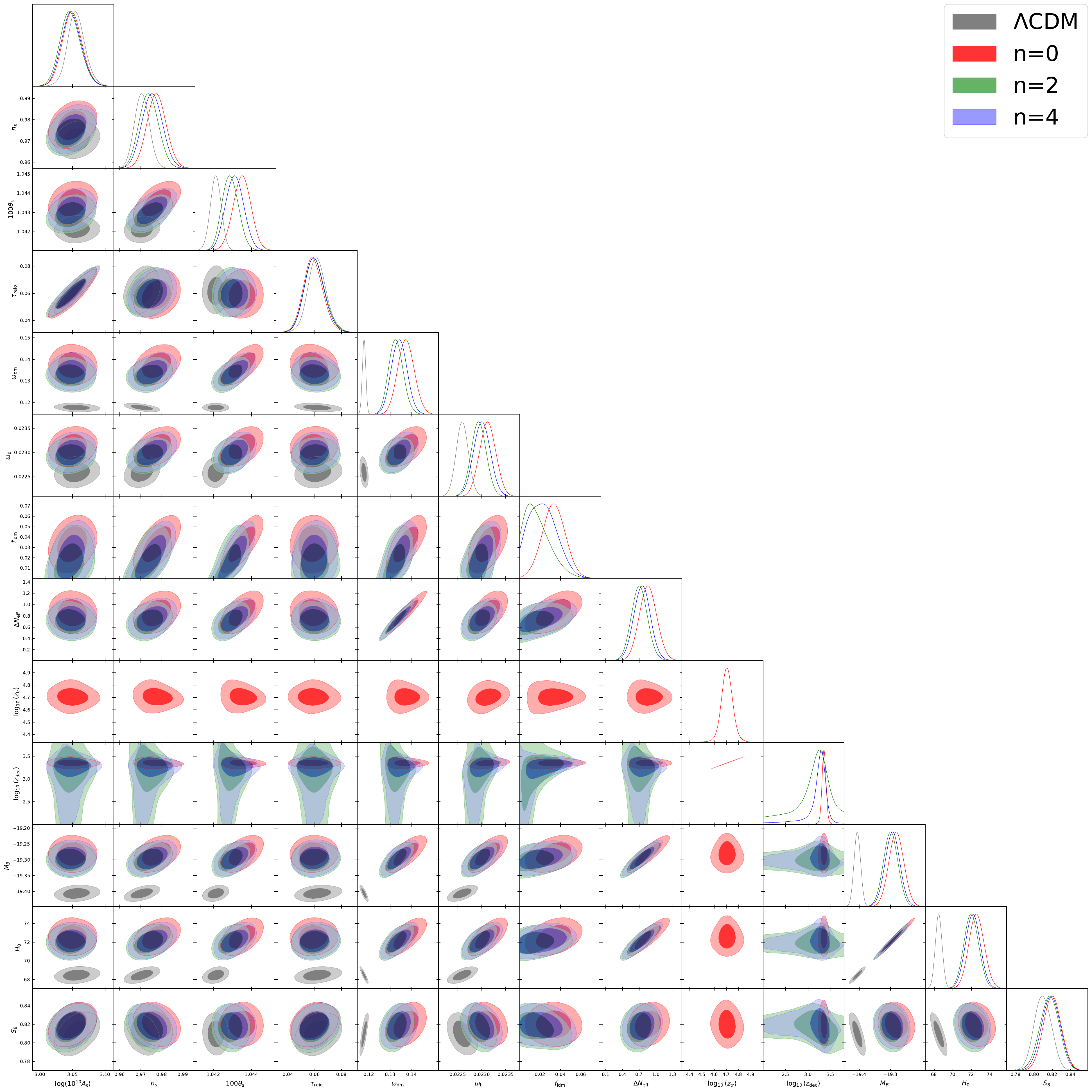}
	\caption{
    1D and 2D posterior distributions of the parameters of the \lcdm (gray), $n=0$ (red), $n=2$ (green), and $n=4$ (blue) models, fitted to the $\mPH$ datasets.
    }
	\label{fig:PH-triangle}
\end{figure}

\section{Sliced Likelihood Analysis of CMB Data}
\label{app:sliced_analysis}

To better understand which multipole ranges and specific CMB spectra drive the statistical preferences and penalties for the \genidm models, we perform a sliced $\Delta\chi^2$ analysis. We focus on the $n=0$ model, which exhibits the strongest preference over \lcdm when local \HO measurements are included, and compare its best-fit points to those of \lcdm. 

For a given multipole range or spectrum, we slice the bandpower data and its corresponding covariance matrix, invert the sub-covariance to obtain a marginal sub-likelihood, and recompute the residual $\chi^2$ for both best-fits. Because this procedure evaluates a marginal $\chi^2$, answering how well each model fits a specific subset of the data on its own without off-diagonal correlations to outside bandpowers, the sliced $\Delta\chi^2$ values do not sum to the total $\Delta\chi^2$ of the full likelihood.

\subsection{Planck Data ($\mPH$ Best-Fits)}

We first examine the Planck high-$\ell$ TTTEEE data using the \texttt{plik\_lite} likelihood, which provides access to the per-bandpower covariance. We use the best-fit points to the $\mPH$ dataset, where the $n=0$ model accommodates a large Hubble constant ($\HO \simeq 72.6~\km/\seg/\Mpc$) and a large amount of dark radiation ($\DNeff \simeq 0.86$). The results are summarized in Table~\ref{tab:sliced_planck}.

\begin{table}[h!]
\centering
\begin{adjustbox}{max width=0.75\columnwidth}
\begin{tabular}{@{}lcccc@{}}
\toprule
\toprule
Range & $n_\text{bins}$ & $\chi^2$(\lcdm) & $\chi^2$($n=0$) & $\Delta\chi^2$ \\ \midrule
All $\ell$ & 613 & 586.34 & 585.38 & $-0.95$ \\
$\ell < 1000$ & 342 & 288.90 & 289.74 & $+0.84$ \\
$1000 \leq \ell < 1500$ & 168 & 188.93 & 187.24 & $-1.70$ \\
$\ell \geq 1500$ & 103 & 105.79 & 107.72 & $+1.93$ \\ \midrule
\textit{Per spectrum (full $\ell$):} & & & & \\
TT & 215 & 209.15 & 209.70 & $+0.56$ \\
TE & 199 & 184.66 & 184.68 & $+0.02$ \\
EE & 199 & 197.83 & 195.65 & $-2.19$ \\
\bottomrule
\bottomrule
\end{tabular}
\end{adjustbox}
\caption{Sliced $\Delta\chi^2$ between the \lcdm and $n=0$ \genidm best-fits to the $\mPH$ dataset, evaluated on the Planck \texttt{plik\_lite} TTTEEE likelihood. A negative $\Delta\chi^2$ indicates a preference for the $n=0$ model.}
\label{tab:sliced_planck}
\end{table}

The analysis reveals a clear physical trade-off. The larger $\DNeff$ required to reduce the sound horizon and raise \HO induces enhanced Silk damping at small scales. The scale-dependent modification of the gravitational potentials caused by the iDM-DR interactions partially compensates for this damping. Table~\ref{tab:sliced_planck} shows that this compensation is highly effective up to $\ell \sim 1500$, where the $n=0$ model actually improves the fit by $\Delta\chi^2 \simeq -1.70$ in the $1000 \leq \ell < 1500$ range. 

However, at $\ell \gtrsim 1500$, the compensation becomes incomplete,
leading to a mild degradation in the high-$\ell$ tail
($\Delta\chi^2 \simeq +1.93$). This degradation is accompanied by
improvements at intermediate multipoles and in the EE polarization spectrum.
For the full Planck high-$\ell$ \texttt{plik\_lite} likelihood, the net
change is small ($\Delta\chi^2 \simeq -0.95$). This illustrates how the
model can support $\HO \simeq 72.6~\km/\seg/\Mpc$ without a significant
penalty from the Planck high-$\ell$ likelihood.

\subsection{ACT DR6 Data ($\mAH$ Best-Fits)}

To understand how the inclusion of ACT DR6 data restricts the viable parameter space, we perform an analogous slicing using the ACT \texttt{MFLike} likelihood. We evaluate the best-fit points to the $\mAH$ dataset, where the $n=0$ model yields $\HO \simeq 70.5~\km/\seg/\Mpc$ and $\DNeff \simeq 0.33$. The results are shown in Table~\ref{tab:sliced_act}.

\begin{table}[h!]
\centering
\begin{adjustbox}{max width=0.75\columnwidth}
\begin{tabular}{@{}lcccc@{}}
\toprule
\toprule
Range & $n_\text{bins}$ & $\chi^2$(\lcdm) & $\chi^2$($n=0$) & $\Delta\chi^2$ \\ \midrule
All $\ell$ & 1651 & 1598.94 & 1604.30 & $+5.36$ \\
$\ell < 1500$ & 462 & 438.66 & 440.17 & $+1.51$ \\
$1500 \leq \ell < 3000$ & 738 & 708.08 & 713.11 & $+5.03$ \\
$3000 \leq \ell < 5000$ & 246 & 227.96 & 226.63 & $-1.34$ \\
$\ell \geq 5000$ & 205 & 223.97 & 224.15 & $+0.18$ \\ \midrule
\textit{Per spectrum (full $\ell$):} & & & & \\
TT & 601 & 565.43 & 566.00 & $+0.57$ \\
TE & 644 & 654.77 & 658.19 & $+3.43$ \\
EE & 406 & 390.40 & 393.10 & $+2.71$ \\
\bottomrule
\bottomrule
\end{tabular}
\end{adjustbox}
\caption{Sliced $\Delta\chi^2$ between the \lcdm and $n=0$ \genidm best-fits to the $\mAH$ dataset, evaluated on the ACT DR6 \texttt{MFLike} likelihood. A positive $\Delta\chi^2$ indicates a preference for \lcdm.}
\label{tab:sliced_act}
\end{table}

The total ACT penalty for the $n=0$ model in the $\mAH$ fit is $\Delta\chi^2 \simeq +5.36$. The slicing demonstrates that the bulk of this constraining power comes from the intermediate-multipole range $1500 \leq \ell < 3000$, which alone accounts for $\Delta\chi^2 \simeq +5.03$. The very-high-$\ell$ ranges ($\ell \geq 3000$) are nearly neutral, indicating that the dominant statistical weight disfavoring the model sits in the $1500 \lesssim \ell \lesssim 3000$ window rather than in the deep damping tail.

Breaking the constraint down by spectrum reveals that the discrepancy is driven almost entirely by the polarization channels, with TE contributing $\Delta\chi^2 \simeq +3.43$ and EE contributing $\Delta\chi^2 \simeq +2.71$, while TT is nearly neutral. This pattern is fully consistent with the physical picture established above: the $1500 \leq \ell < 3000$ range is precisely where the scale-dependent offset of Silk damping by the DM-DR interactions becomes incomplete. Because ACT possesses high precision in this specific polarization window where Planck does not, it catches the residual damping, thereby tightening the constraints on \DNeff and limiting the maximum \HO reachable by the model.

\clearpage

\bibliographystyle{JHEP}
\bibliography{refs}

\end{document}